\documentclass{article}
\usepackage[scaled]{helvet}

\usepackage[english]{babel}

\usepackage[letterpaper,top=2cm,bottom=2cm,left=3cm,right=3cm,marginparwidth=1.75cm]{geometry}

\usepackage{amsmath}
\usepackage{graphicx}
\usepackage[dvipsnames]{xcolor}
\usepackage[colorlinks=true, allcolors=teal]{hyperref}
\usepackage{xspace}
\usepackage{csquotes}
\usepackage{setspace}
\usepackage{tcolorbox}
\usepackage[round]{natbib}
\newcommand{\kmer}{$k$-mer\xspace}
\newcommand{\kmers}{$k$-mers\xspace}
\title{Advances in practical \emph{k}-mer sets: essentials for the curious}

\begin{document}
\author{Camille Marchet $^1$ \\$^1$ UMR9189 CRIStAL, Univ Lille, CNRS, Centrale, F-59000 Lille, France\\camille.marchet@univ-lille.fr}
\date{}
\maketitle
 \begin{abstract}
  This paper provides a comprehensive survey of data structures for representing \emph{k}-mer sets, which are fundamental in high-throughput sequencing analysis. It categorizes the methods into two main strategies: those using fingerprinting and hashing for compact storage, and those leveraging lexicographic properties for efficient representation. The paper reviews key operations supported by these structures, such as membership queries and dynamic updates, and highlights recent advancements in memory efficiency and query speed. A companion paper explores colored \emph{k}-mer sets, which extend these concepts to integrate multiple datasets or genomes.
 \end{abstract}
%

\maketitle
\onehalfspacing
\section{Introduction}

In the era of high-throughput sequencing, string algorithms are indispensable tools for the analysis of biological data. Sequencing technologies generate massive amounts of data by extracting numerous reads—short substrings of DNA or RNA from biological samples. These reads, which range in length from 50 to several thousand characters, are often subject to errors, making the analysis of sequencing data a complex problem. Central to this analysis is the task of string matching, where short, fixed-length substrings known as \kmers, are identified and analyzed. Over the past decade, \kmer-based methods have gained significant popularity due to their scalability and simplicity. These methods have been successfully applied across various biological domains, including genome [\cite{bankevich2012spades}] and transcriptome assembly [\cite{bushmanova2019rnaspades}], transcript expression quantification [\cite{patro2015salmon}], metagenomic classification [\cite{wood2014kraken}], and genotyping [\cite{iqbal2012novo,krannich2022population}]. Emerging applications include antibiotic resistance surveillance and detection [\cite{marini2022towards, bonin2023megares}], and the curation  \kmer  signatures catalogs for cancer or other illnesses [\cite{nguyen2021reference,riquier2021kmerator}]. 

A key challenge in \kmer-based methods is the efficient storage and querying of the vast sets of \kmers  generated from sequencing data. As datasets grow in size and complexity, minimizing the storage requirements and query times for \kmer sets has become a crucial area of research.
A previous work [\cite{chikhi2021data}] provides a complete mathematical analysis of \kmer sets with space and time lower bounds. In this paper, I present a more broadly accessible survey of data structures designed for indexing \kmers, focusing on data-structures that have been used in practice. I categorize these structures and draw connections between them.
I intentionally omit detailed discussions of their applications to specific biological problems, as this is the purpose of a companion article [\cite{Marchet2024colored}]. Instead, I provide a high-level overview of the data structures themselves, offering insights into their underlying principles and practical implementations.

\section{Preliminaries}
\subsection{De Bruijn graphs}
\begin{figure}[ht]
    \centering
    \includegraphics[width=\textwidth]{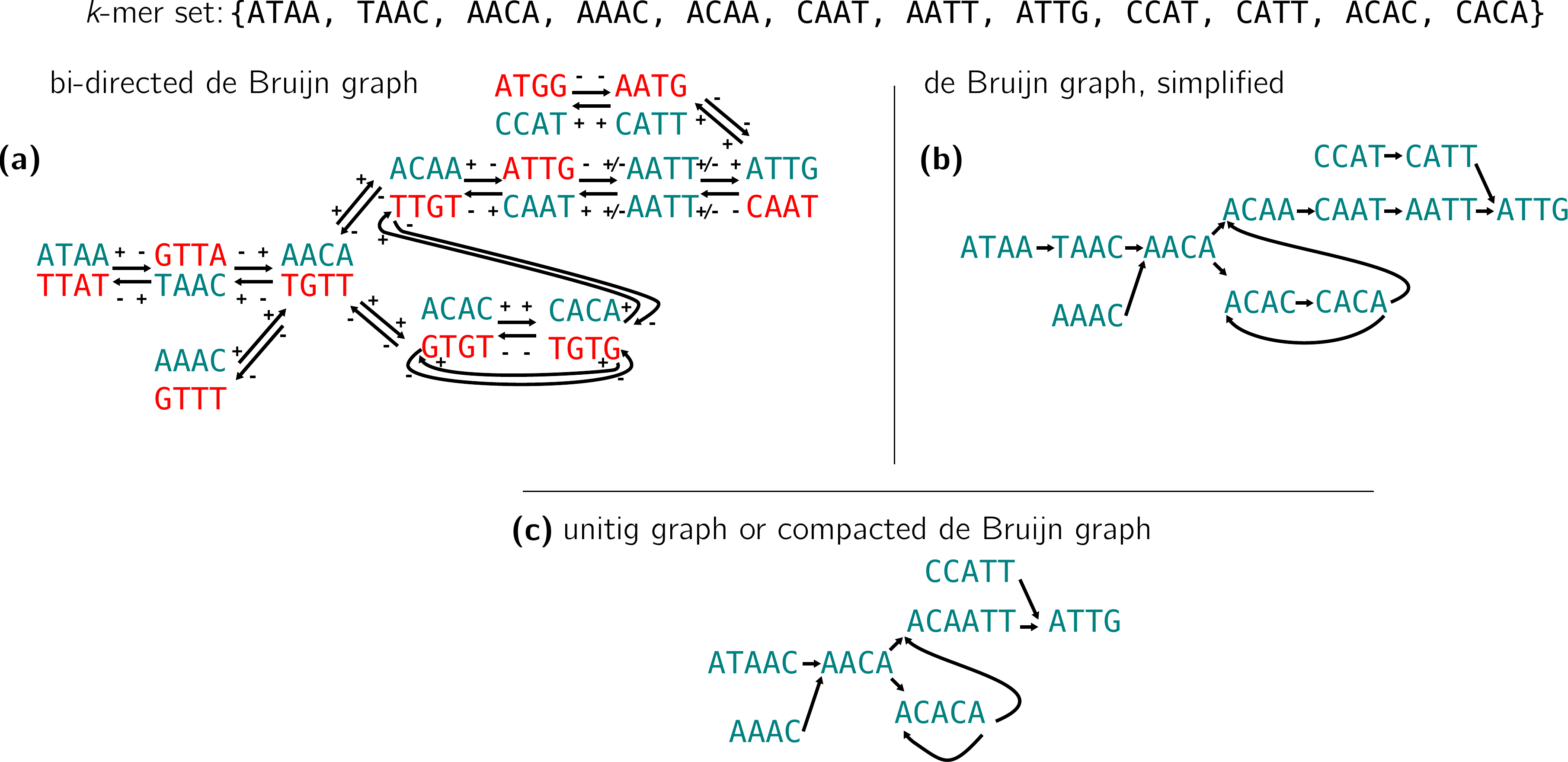}
    \caption{Top of the figure: the \kmer set that is represented by the de Bruijn graphs. (a) A bi-directed de Bruijn graph. Forward \kmers are in blue, reverse in red. (+,+) egdes indicate an overlap between forward and forward \kmers, (-,-) are edges between reverse and reverse \kmers. (+,-) and (-,+) denote forward-reverse and reverse-forward nodes. I draw the reader's attention on the +/- symbols on some edges. This happens because they connect to and from a \kmer whose forward and reverse complement sequences are the same. This is one of the numerous edge case that must be taken care of when implementing de Bruijn graphs. (b) The simplified version that shows only one occurrence in the (forward, reverse) pairs, that is used in examples of this manuscripts. (c) The same graph but whose \kmers have been compacted in a unitig graph.
    }
    \label{fig:dbg_intro}
\end{figure}

In this paper, we consider a common definition of the de Bruijn graph in which distinct \kmers (words of size k) extracted from the sequences to be indexed are nodes (node-centric definition).
More precisely, \kmers and their reverse complements are represented by the same node (the graph is called bi-directed, see Figure~\ref{fig:dbg_intro}).
This bi-directed definition is particularly convenient to build graphs from the general case of non-stranded sequencing data.
The convention that the smallest \kmer (e.g. lexicographically) is considered the forward \kmer and the other is its reverse complement. Directed edges are drawn between nodes sharing k-1 exact overlaps on their forward or reverse representation. 
Other definitions of de Bruijn graph exists, with forward and reverse \kmer separated, or \kmers on edges~[\cite{rahman2022assembler}]. 

De Bruijn graphs are a scalable tool adapted for assembly since the advent of high-throughput short reads. They offer at most four possibilities (A, C, G, T) as the next nucleotide when assembling a contig, making graph construction and traversal feasible even for high coverages with respect to the size of genomes. But aside from assembly, de Bruijn graphs have general properties that are relevant for indexing datasets and genomes. 

First, any set of distinct \kmers implicitly forms a de Bruijn graph since k-1 overlaps can be deduced from \kmers. De Bruijn graphs preserve the structure of the original sequence to some extent (hence facilitating genome and transcriptome assembly), which makes them highlight recognizable patterns of variants and enable error filtering through \kmer abundance and specific topological patterns. I recommend reading more \kmer key techniques reviewed in \cite{jenike2024guide}, with a focus on using \kmer set profiles for genome analysis.


Additionally, de Bruijn graphs allow the construction of shorter strings than simple concatenations of \kmers, which can be used to encode \kmer sets with fewer bits (reviewed in subsection~\ref{section:SPSS}). 
Perhaps the most well-known example is unitigs. A unitig is a stretch of DNA that is supposed to be unambiguous in the graph - meaning there are no branches in the path when connecting one \kmer to the next (see (c) in Figure \ref{fig:dbg_intro}). Compare for instance the number of symbols needed for the unitig \texttt{ACAATT} to the concatenation of its \kmers: \texttt{ACAA|CAAT|AATT}.

In the following, I review the different techniques practically utilised to represents sets of \kmers, or de Bruijn graphs for a single dataset. I spend much tome on \kmer tools whose main purpose is to count \kmers, such as KMC2/3~[\cite{deorowicz2015kmc,kokot2017kmc}] or Jellyfish~[\cite{marcais2012jellyfish}]. They are quite often used as preliminary method for  set representations reviewed here. I also omit some of the methods developed and embedded within genome assemblers and focus on standalone tools. Colored de Bruijn graphs, introduced in \cite{iqbal2012novo}, integrate multiple genomes or samples into a single structure, factorizing common parts and highlighting differences. Because they justify their own developments, I review them in a companion paper [\cite{Marchet2024colored}].

\section{Overview of \emph{k}-mer sets}

\begin{figure}[ht]
    \centering
    \includegraphics[width=1\textwidth]{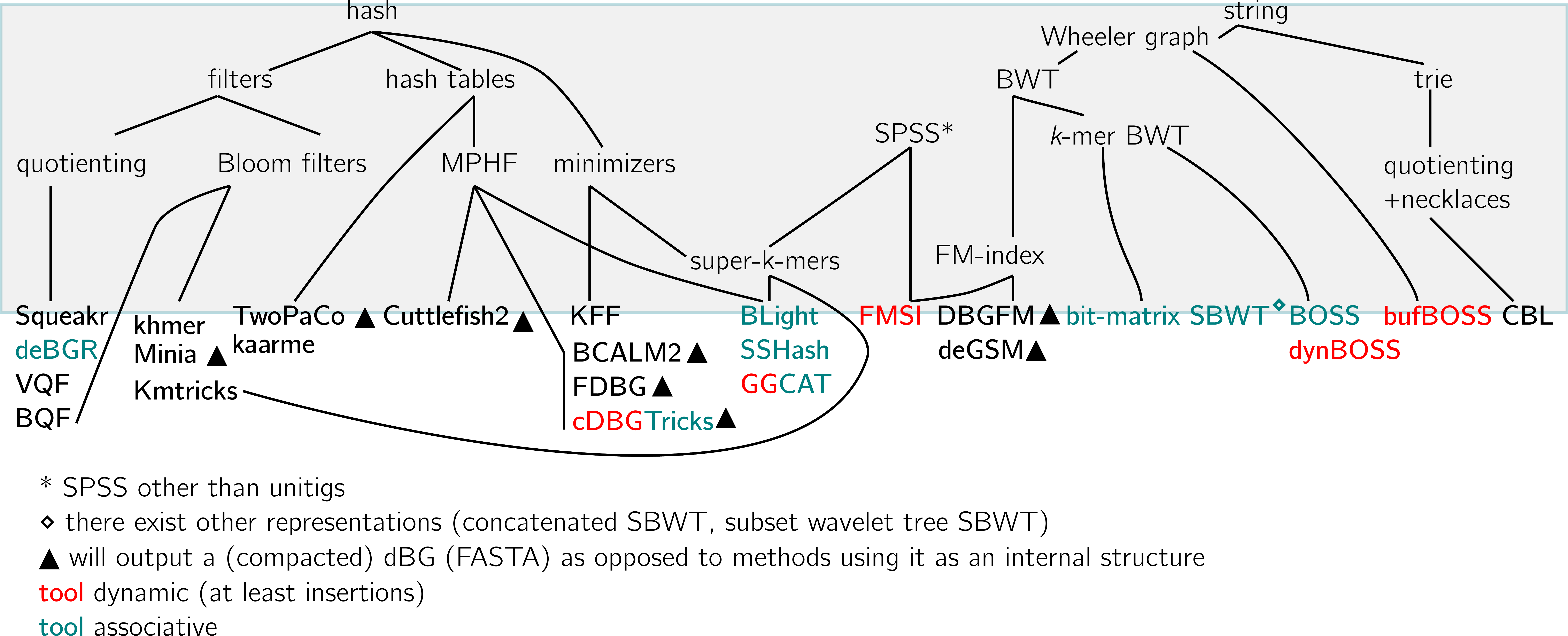}
    \caption{Landscape of \kmer sets, starting from internal representation of \kmers based on strings or hashes. Tools to build (compacted) de Bruijn graphs (triangle) build the graph or unitigs from an input \kmer set and frequently output them in a FASTA format. $K$-mer sets allowing insertions (red), and sometimes deletions and set operations on an input \kmer set. Associative structures (blue) allow to pair \kmers with pieces of information, three of them are information specific: BQF, deBGR and Squeakr associate \kmers to count. The others are generalist \kmer dictionaries.}
    \label{fig:genealogy0}
\end{figure}

The plethora of methods for representing \kmer sets, each with slight variations, can be overwhelming for newcomers or researchers looking to utilize these methods in their analyses. Many tools incorporate other tools in their algorithms or implementations, creating a nested structure that complicates navigation.
The goal of this section is to provide clarity by categorizing these methods to some extent, recognizing that such categorizations may not perfectly reflect reality but can help structure a coherent view (Figure \ref{fig:genealogy0}). Key observations include that some methods make direct use of \kmer seen as lexicographic units, while others rely on \kmers as integers.

\subsection{Operations}
\paragraph{\textbf{Membership queries}} A fundamental operation is checking the presence or absence of a given \kmer in the indexed collection. This is supported by all methods, with varying query times. Filter-based approaches can have false positives (thus called approximate membership queries), while other methods provide exact membership queries.\\
In terms of performance, the reviewed methods exhibit a range of time and space trade-offs. 

Cache locality is often part of the discussion in modern methods, as it is crucial for optimizing program performance, especially when working with large datasets. It aims at maximizing the probability that for a given set of bits, the likely next accessed bits are close in the processor's cache. This minimizes delays and maximizes processing speed. Methods that have better cache locality often outperform others in practical scenarios, in particular when consecutive \kmers are queried (sometimes called batch queries, as opposed to single queries).

\paragraph{\textbf{Navigation}} Some structures also enable navigating the de Bruijn graph induced by the \kmer set, allowing to find adjacent \kmers and traverse the graph. This is particularly useful for assembly or variant detection applications. According to the structure design, navigational queries can be cheaper than membership queries.
\paragraph{\textbf{Ranking}} Some methods support ranking \kmers, for example to find the \kmer at a given position in the lexicographic ordering of the \kmer set.
\paragraph{\textbf{Set operations}} Operations like set union, intersection, and difference have been implemented in a few recent methods, enabling complex queries across multiple datasets.
\paragraph{\textbf{Dynamic updates}} While most methods are static, requiring a full rebuild upon changes in the \kmer set, a few dynamic structures allow efficient insertion and, more rarely, deletion of \kmers.
\paragraph{\textbf{Count \emph{k}-mers}} Methods have been developped especially for this task, as this is an essential pre-processing to many pipelines. Counts can be approximate or exact.

\subsection{Landscape of practical \emph{k}-mer sets}

We know the worst-case lower bound for representing a set of \kmers in a membership structure, under the assumption that \kmers are drawn from a uniform distribution, given by information theory~[\cite{conway2011succinct}]: $\#\text{bits} = \log_2 \binom{4^k}{n}$, with $n$ the number of \kmers in the set. Representing all canonical 31-mers from a human genome would require approximately 15 GB. We will see that this lower bound tells actually not much, as it is often beaten by specialized structures. Another bound has been set of structures that allow only navigational queries can go lower, down to 3.24 bits/\kmer [\cite{chikhi2014representation}].

Strategies leveraging lexicographic context or redundancy of consecutive \kmers (reviewed in section~\ref{section:strings}) try to make the most of the data specificity: \kmers are extracted from a structured, non-random genomic context, sequencing data can contain a lot of redundancy and repeats.

\begin{tcolorbox}[colback=lightgray!30!white, colframe=lightgray!50!darkgray, sharp corners, boxrule=0mm, title=Leverage string properties: example]
For instance, consider the string \texttt{ACTGAGCTGAGCTGA}, which contains 7 5-mers. I chose it voluntarily redundant for the sake of the example. Applying a lexicographic transform on the string can change it into  \texttt{AG\$GGGGATTTAACCC} (a dummy \texttt{\$} exists for technical reasons). \texttt{AG\$GGGGATTTAACCC} has convenient runs of characters that can be further compressed, intuitively: \texttt{AG\$(G,4)A(T,3)(A,2)(C,3)}.\\

Another way to reach a more compact representations uses the \kmer redundancy, as in the former unitig construction example: \texttt{ACAA|CAAT|AATT} $\rightarrow$ \texttt{ACAATT}.
\end{tcolorbox}

In the second case, integer fingerprints can be stored instead of \kmers (reviewed in section \ref{section:hash}). Fingerprints refer to compact, integer representations of \kmers. When a \kmer is added to the structure, its fingerprint (often a hash) is stored instead of the full \kmer. Lossy fingerprints (carrying less information than needed to retrieve the original \kmer) are one source of false positives, but represent a gain in bits per \kmer in the structure. 

\begin{tcolorbox}[colback=lightgray!30!white, colframe=lightgray!50!darkgray, sharp corners, boxrule=0mm, title=Fingerprints: example]
Consider the \kmer \texttt{ATGGC}. The most straightforward, lossless, fingerprint is its binary encoding using the property that we can write \texttt{A $\rightarrow$ 00, C $\rightarrow$ 01, G $\rightarrow$ 10, T $\rightarrow$ 11}. It is therefore encoded on 2$\times$5 bits as \texttt{0011101001}, or as the integer \texttt{489}.
Consider the \kmer \texttt{AGGGC}, we obtain \texttt{0010101001} with an expected redundancy in the notation.
\\

Fingerprints based on hashing have different properties and aim at 1-uniformity across the integer space and 2-determinism. \texttt{ATGGC} and \texttt{AGGGC} will likely be associated to very different values. We can also use lossy fingerprints to save space, for instance on 8 bits, and obtain for instance \texttt{ATGGC $\rightarrow$ 00100001} and \texttt{AGGGC $\rightarrow$ 01100000}.

\end{tcolorbox}

Associativity (creating structures associating \kmer/value) can be achieved below the lower bound, costs fluctuate with input data structure, relying on the dataset's characteristics to optimize space. They rely on dictionary structure (key,value pairs structures with the \kmer as keys), or on structures that can associate \kmers using their ranks.

Another source of difference in the methods is that many methods are static, computed on a specific set and requiring a complete rebuild for any changes, a strategy achieving extremely low memory footprints. The possibility to add or remove \kmers without false positives is still rare, but available~[\cite{martayan2024conway,sladky2024function,hannoush2024cdbgtricks,alipanahi2021succinct,alanko2021buffering}], albeit at higher costs than the most recent structures. In a different approach, some methods allow updates at the price of a proportion of false positives when queried, which can be manageable with large datasets (reviewed in subsection~\ref{section:filters}).

Query performance is another critical aspect, that can benefit too from observations on the data specificity, notably by fitting the computer's cache size with \kmers that are likely to be queried altogether. The quickest methods, e.g. [\cite{pibiri2022sparse}], query \kmers in a few hundreds of nanoseconds.

\section{\emph{K}-mers as strings}\label{section:strings}
The methods reviewed in this subsection try and find some types of character redundancies in the \kmers in order to represent the \kmer space using less space.

\subsection{Spectrum preserving string sets (SPSS)}\label{section:SPSS}

For what matters in this article, \emph{spectrum preserving string sets} have a confusing name. The goal of the techniques I review here is to encode a \kmer set within a string set. Indeed, methods constructing these sets usually do not preserve spectrum (\kmers with frequency), but only the \kmer set. I'd rather call them \emph{set preserving string sets}\footnote{as coined by A.L.} if I had a choice.

\begin{figure}[ht]
    \centering
    \includegraphics[width=0.8\textwidth]{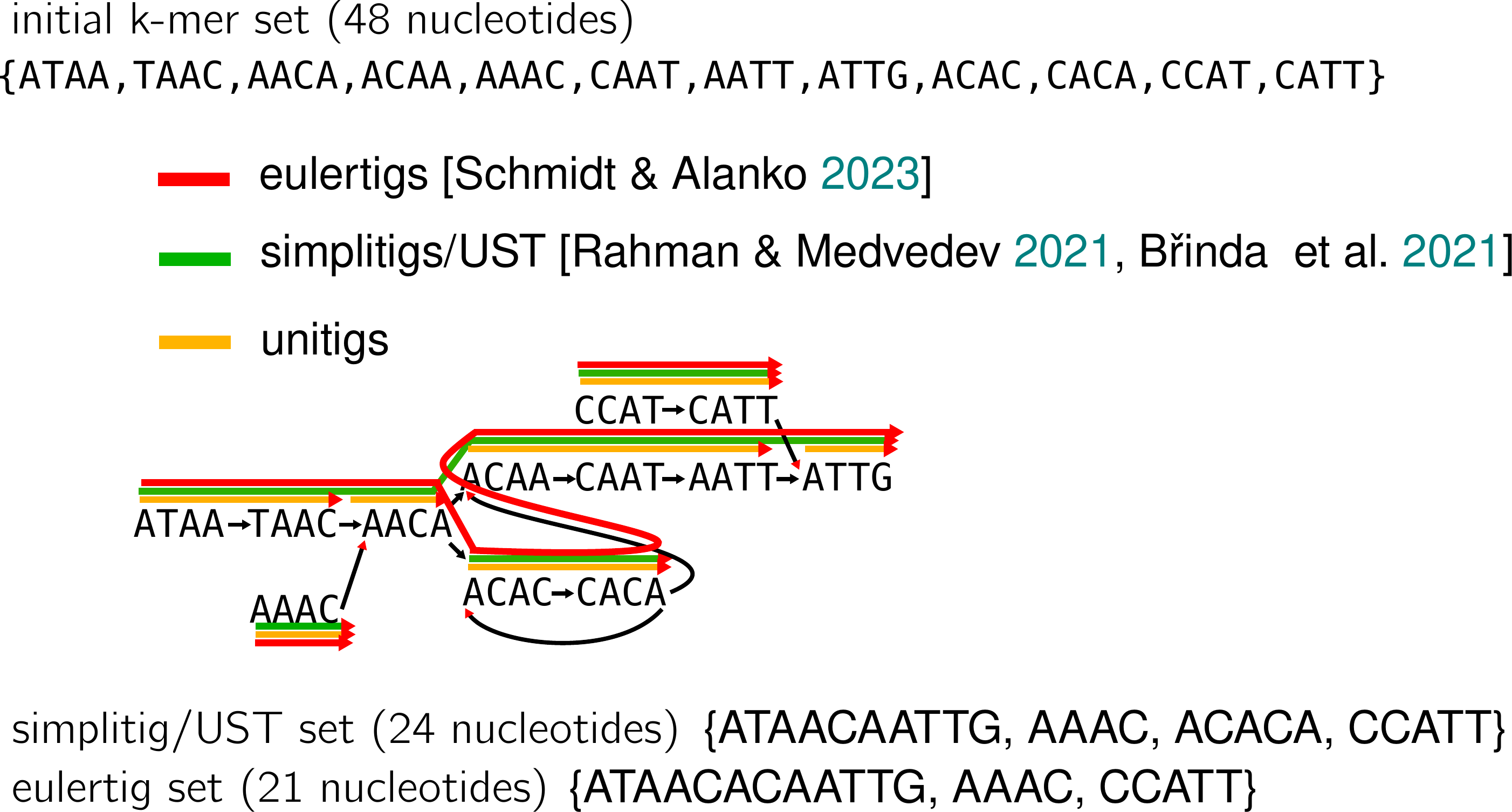}
    \caption{Different examples of SPSS built from a same set of \kmers.}
    \label{fig:spss}
\end{figure}

De Bruijn graphs allow constructing strings shorter than concatenated \kmers, enabling more compact \kmer set encoding. These string sets, have been formalized for \kmers by \cite{rahman2021representation}, although they appeared earlier in B\v rinda's work~[\cite{brinda2016nouvelles}]. 
Given a \kmer input, a spectrum preserving string set is a plain text representation from which all initial \kmers and nothing else can be spelled (Figure \ref{fig:spss}).

Paths in the de Bruijn graph can represent longer strings. Finding an SPSS involves identifying paths covering all graph nodes and processing nodes within each path with a compaction algorithm. 

\begin{tcolorbox}[colback=lightgray!30!white, colframe=lightgray!50!darkgray, sharp corners, boxrule=0mm, title=Compaction: example]
Intuitively, compaction is merging consecutive strings as follows: if one \kmer ends spells "ATG" and the next \kmer is "TGA", they can be merged into a longer sequence composed of the first \kmer and the last base of the second, "ATGA", because they overlap.
\end{tcolorbox}

An optimal solution for the SPSS problem does compactions so it minimizes the number of strings while ensuring each \kmer appears only once in the string set.

Unitigs, the product of spelling paths with no branch in the de Bruijn graph, are non-optimal SPSS, often used for more compact de Bruijn graph representations (unitig graphs are called compacted de Bruijn graphs), with different algorithms and trade-offs, including  BCALM2~[\cite{bcalm2}], TwoPaCo~[\cite{10.1093/bioinformatics/btw609}], Cuttlefish2~[\cite{khan2022scalable}], and GGCAT~[\cite{cracco2023extremely}]. Unitigs remain widespread because they are balanced options between shortest sequences and biological meaningfulness, they also are a intermediary component of short read assemblers. Moreover, they are often constructed after a \kmer filtering and error correction pass on the initial graph, and provide a cleaner set.

Recent heuristics, such as simplitigs~[\cite{bvrinda2021simplitigs}] and USTs~[\cite{rahman2021representation}], come close to optimal solutions for the SPSS problem. They allow to store 31-mers from a human genome in 2 GB or less. Eulertigs~[\cite{schmidt2023eulertigs}] further provide a linear algorithm for optimal solutions based on Eulerian cycles.

The most recent contribution, masked superstrings~[\cite{sladky2023masked}], unify different proposals into a common framework. It generalizes the compactions to overlaps smaller than k-1, which creates spurious \kmer but increases the capacity to elongate strings. They use masks (binary arrays) to indicate which parts of the superstring correspond to the \kmers of interest. The GGCAT method for constructing de Bruijn graphs also proposes different options for SPSS. Before that, a different route was taken by matchtigs, that allow to re-use a \kmer several times to achieve larger superstrings [\cite{schmidt2021matchtigs}]. 

While SPSS can represent \kmers compactly, they are not directly easy to query. They are combined with other techniques to create \kmer structures, for instance, in the FMSI~[\cite{sladky2024function}] index, the KFF tool [\cite{dufresne2022k}] and coupled to hashing for membership queries for matchtigs [\cite{schmidt2021matchtigs}].\\

\paragraph{\textbf{Example use cases of SPSS}}
SPSS are not usually used per se in pipelines, but rather as a piece of larger algorithms. For instance, recently, SPSS (USTs) were used within a compression algorithm that reduces \kmers and associated counts footprint [\cite{Rossignolo2024}]. As we will see later in the article (subsection \ref{section:hashtables}), they are a key component for modern \kmer hash tables.

\subsection{BWT-based methods}
The Burrows-Wheeler Transform (BWT) is a data transformation technique used primarily in data compression. It makes use of recurrent neighborhoods surrounding characters in a string to rearrange the string into runs of similar characters. This makes the string more amenable to compression. It is then possible to access any origin substring by querying the transform with a small quantity of auxiliary information in comparison to the original string size.
An alternative strategy for \kmer sets is therefore using methods based on the BWT to index \kmer sets. These methods operate directly on sequences, grouping similar segments for compression. Among first proposals, many relied on an index based on the BWT, the FM-index [\cite{ferragina2005fmindex}]. However, the BWT excels on large texts, not on sets of words, and its performances decline when there are errors in the dataset. That is why early uses of BWT for indexing \kmer sets have employed longer strings simplifying the \kmer sets, such as unitigs [\cite{chikhi2014representation}].

\begin{figure}[ht]
    \centering
    \includegraphics[width=0.6\textwidth]{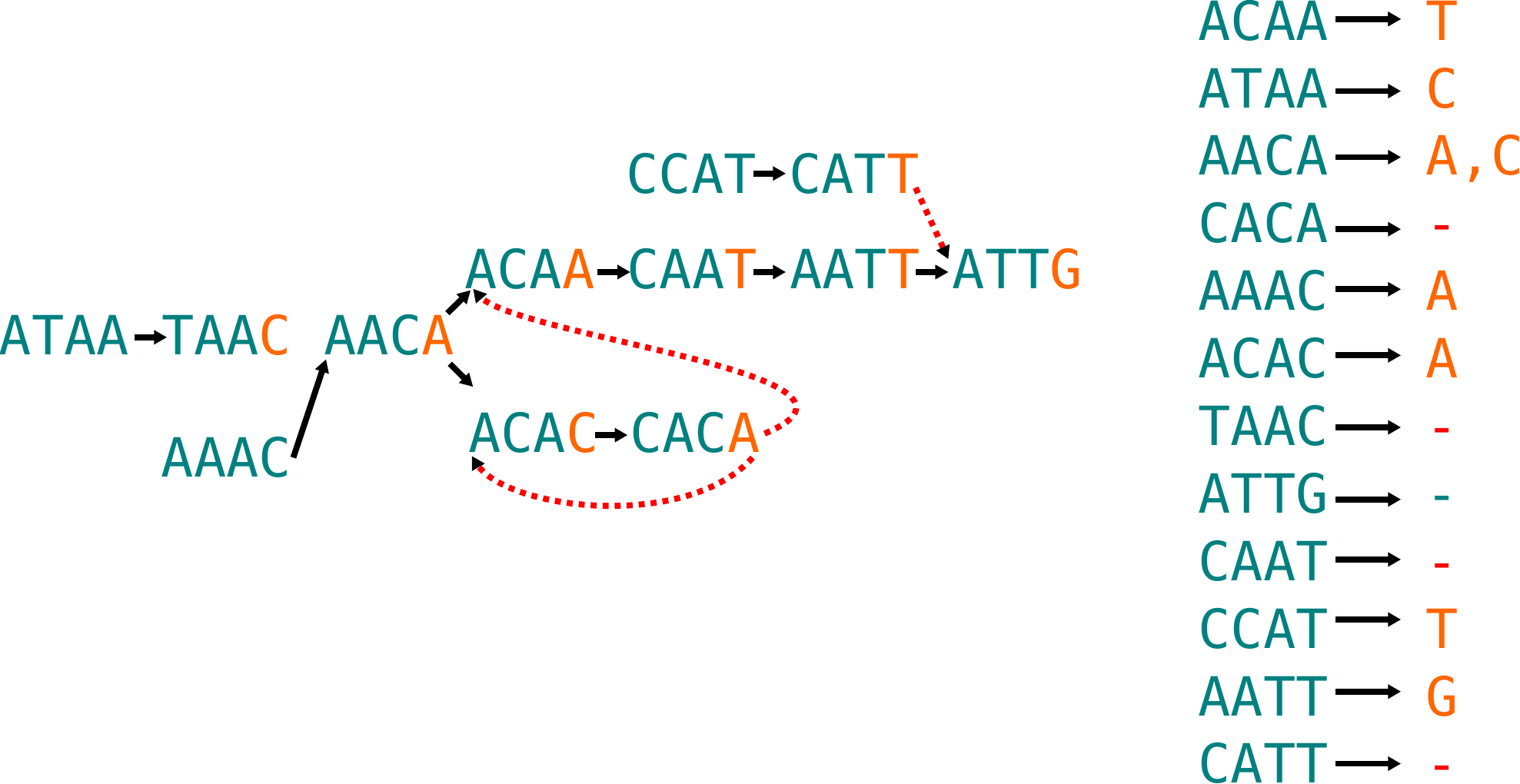}
    \caption{Intuition of the SBWT of a de Bruijn graph. The SBWT retains connections between \kmers, and leaves some edges (dashed in red) when the \kmer already has another recorded parent. Last nucleotide of children \kmers for retained edges are associated to \kmers sorted colexicographically. The rightmost colored vector is in essence what is stored in the bit-matrix SBWT and allows retrieve \kmers. In practice, some more dummy nodes are needed and specific data structures are needed to retrieve efficiently the information.}
    \label{fig:sbwt}
\end{figure}

The BOSS structure~[\cite{bowe2012succinct}] specialized on \kmer sets and introduced a representation for edge-centric de Bruijn graphs (a definition where \kmers are on edges).
The BOSS structure was also turned into a dynamic structure allowing insertions and deletions in \kmer sets, through a system of buffers and sporadic rebuild of the structure~[\cite{alipanahi2021succinct}]. Another dynamic structure, called BufBOSS~[\cite{alanko2021buffering}] uses similar principles but despite its name, is not based on BOSS. Instead, it uses an efficient string indexing, similar to the BWT, but in graph form (the Wheeler graph).\\
Another advancement is the SBWT~[\cite{alanko2023small}] (Figure \ref{fig:sbwt}). The SBWT is also a BWT-like transform, offering node-centric representations of a de Bruijn graph, and enhancing query efficiency. 

Similar to the BWT being coupled with other structures to create a index on the strings, the SBWT can be coupled to data-structures to create an index on \kmers.
Therefore, all BOSS, SBWT, \ldots implementations have a range of time-space tradeoffs depending on the underlying data structures and on the compression rate. One variant of SBWT is called the bit-matrix SBTW. The bit-matrix SBWT can index distinct 31-mers of the human genome in approximately 2 GB, and achieved competitive query speed when equipped with auxiliary query structures~[\cite{alanko2023longest}], for the price of a larger index. The BOSS and bit-matrix SBWT can be associative structures by using the colexicographic rank of \kmers as an index.\\

\paragraph{\textbf{Example use cases of BWT-based methods}}
Burrows-Wheeler Transform based indexes have become fundamental tools in bioinformatics due to their ability to efficiently compress and search genomic data. BWT is used in countless applications, notably in read alignment tools. In the context of \kmer indexing, 
the BOSS structure was implied in different works dealing with surveillance of foodborne pathogens. It was integrated in a reference-free metagenomics SNP caller [\cite{alipanahi2020metagenome}], whose advantage is to be able to extract complex embedded SNPs, as in different samples of a beef production system for detecting antibiotic resistance genes. Similarly, an index based on the bit-matrix SBWT was embedded a in pipeline for genomic epidemiology dealing with mixed samples of a target pathogen [\cite{maklin2021bacterial}]. 
Recently, as an associative array for \kmers based on the SBWT was used to preprocess data for a machine learning model to classify antimicrobial resistance (AMR) in Mycobacterium tuberculosis genomes [\cite{10.1093/bioinformatics/btae243}].

\subsection{Tries and variations}

A trie is a tree-like data structure used to store a dynamic set of strings, where each node represents a single character of a string. Tries store strings in a way that allows common prefixes to be shared among different strings. Each path from the root to a leaf node in the trie represents a complete string. Tries are not too common for \kmer indexing, but appear from time to time in the literature~[\cite{agret2020redoak,holley2016bftrie}].

~\cite{martayan2024conway} noticed that the necklace text transform [\cite{sawada2017practical}] increases the number of shared prefixes in a \kmer set. The transform is related to the BWT as it is also based on lexicographic rotations of words. Instead of a regular trie, the CBL structure is divided in two: prefixes and suffixes. Its stores once common prefixes to save space, and the rank of prefixes associates them with suffixes. 
The necklace transform also tends to improve cache efficiency, as consecutive \kmers will share prefixes.

The strategy where \kmers are split in a left (prefix) and right (suffix) part is also called quotienting, and more frequently used in hash-based techniques (see subsection~\ref{section:hashtables}).

CBL uses 200 GB to represent 31-mers of the human genome, trading space for full dynamicity. It is one of a few, with FMSI, that permits set operations on the \kmers of the structure: intersection, union, difference.\\

\paragraph{\textbf{Example use cases of tries}}
CBL is used in the query builder Grimr [\cite{Ingels2024}] that helps querying multiple samples based on meta-data constrained. Another trie structure was used to detect gene signature in a rice pangenome [\cite{agret2020redoak}].

\section{\emph{K}-mers as hashes}\label{section:hash}
The simplest \kmer integer representation is the succinct indicator bitmap used in \cite{conway2011succinct}. It uses the integer representation of \kmers to address them in a bit array, and can be compressed in a space close to the information-theoretic minimum while still allowing efficient access.\\

Other methods apply hash functions to \kmers first. These methods aim at populating tables with \kmers but have to deal with an initial skewed distribution, due to the repetitive and non-uniform content of \kmer sets with respect to the whole $4^k$ universe.
Using directly the integer representation of the \kmers from these skewed distribution could lead to cluttered regions in tables, with impacts on space, insertion time and query efficiency. The reviewed methods rely on hashing, transforming \kmers into integers via hash functions, to uniformly fill the allocated table or bit set.

\subsection{Structures with false positives: filters}
\begin{figure}[ht]
    \centering
    \includegraphics[width=\textwidth]{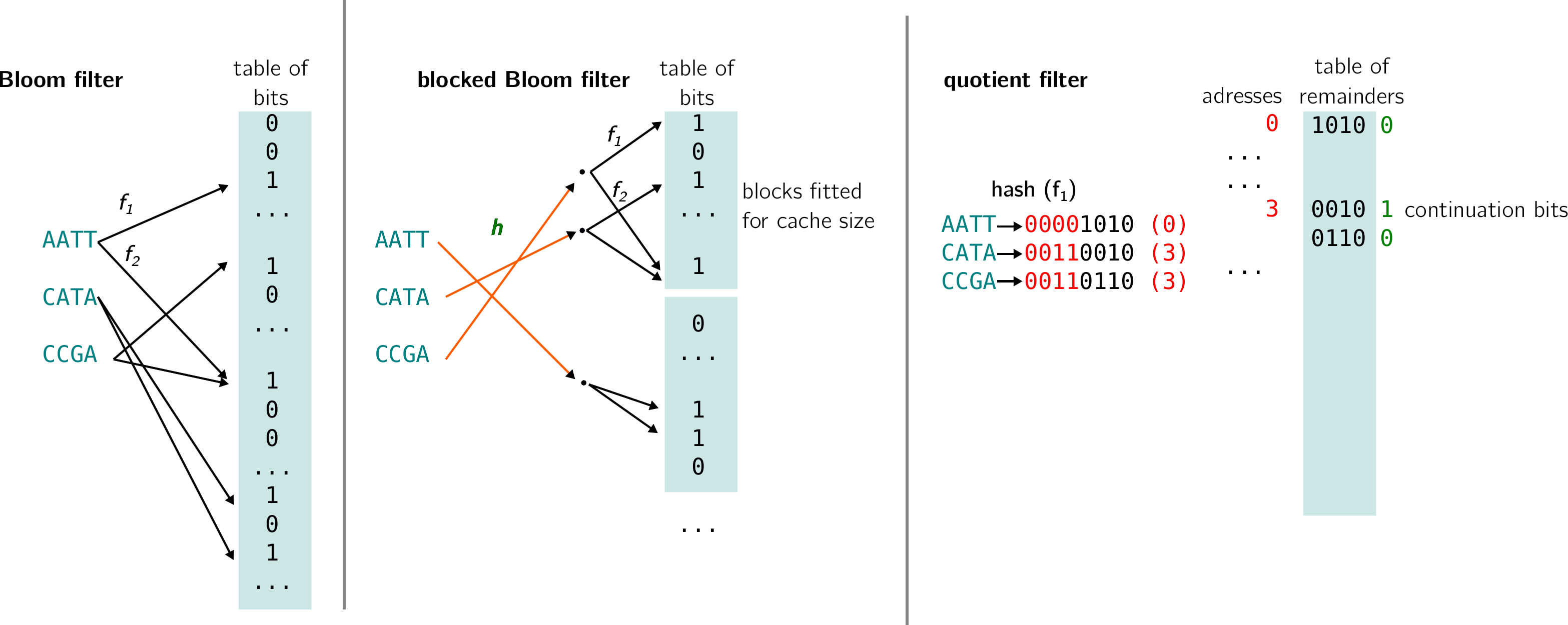}
    \caption{Intuition of \kmer encoding with filters. $K$-mer are hashed for addressing. Left: in the Bloom filters, bits are set to 1 for the given addresses, with possible collisions (e.g. ). Middle: the blocked Bloom filter divides a Bloom filter in blocks, where \kmers are addressed with a first hash function. Then blocks are filled as standard Bloom filters. Right: In quotient filter, the quotient (red part) of the hash addresses the remainder, and a probing strategy takes place when several remainders end up at the same location. In that example it is the case for CATA and CCGA here, a continuation bit indicates that there are several remainders to be checked. I chose to represent two hash functions for the Bloom filter, as in practice many tools use 1 to 3.}
    \label{fig:amq}
\end{figure}
\subsubsection{Bloom filters}\label{section:filters}
Bloom filters are probably the most common probabilistic (yielding false positive) structure for \kmer sets. 

Bloom filters are a space-efficient data structure that use a bit vector and multiple hash functions to represent a set of \kmers. During construction, each \kmer is hashed, and the corresponding bits in the bit vector are set to 1. To check if a \kmer is in the set, the same hash functions are applied, and if all corresponding bits are 1, the \kmer is considered to be present. However, collisions (different \kmers hashing to the same bits) can lead to false positives (this type of structure is sometimes called \emph{approximate membership query} or AMQ) (see Figure \ref{fig:amq} left).


Bloom filters are used because the theoretical space that has to be reserved to store any possible \kmer ($4^k$) using a bit set far exceeds practical datasets. For example, a set of 10 billion \kmers (even in large metagenomics datasets) is orders of magnitude smaller than $4^{31}$. Therefore, they aim to project \kmers into a constrained space, hashing them into integers within a range (e.g., $0$ to $2^{32}-1$). They use hashing to uniformly distribute \kmers, avoiding dense and sparse regions in the filter. Using Bloom filters and a high rate of false positives ($>0.1\%$) can lead to representing 31-mers of the human genome below 5 GB.


Bloom filters are straightforward to implement and, to some extent, dynamic. However, over time, the bit array can become saturated with 1s, necessitating resizing. The query is also not as fast as other methods based on hashing. When $k$-mers are the input, Bloom filters can benefit from highly efficient construction with Kmtricks~\cite{10.1093/bioadv/vbac029}.\\
Some works aimed to restore locality properties lost in hashing, such as blocked Bloom filters~[\cite{putze2010cache}] or interleaved Bloom filters~[\cite{dadi2018dream}]. Blocked Bloom filters place \kmers in in blocks within a bit array, and these blocks fit the cache (typically on 64 bits), so queried \kmers require loading a single location. This strategy improves the query but require $\sim$ 30\% more space (Figure \ref{fig:amq}).

There exist variations to Bloom filters, such as the counting Bloom filter that registers approximate counts instead of presence/absence [\cite{Fan2000}]. Related to our survey, recent improvement involving Bloom filters aim at representing abundances associated to \kmers in a compressed way~[\cite{shibuya2022space}]. Other works targeted the reduction of false positives in Bloom filters by indexing storing the list of s-mers $(s<k)$ composing a \kmer instead of the \kmer itself~[\cite{robidou2021findere,robidou2023fimpera}].

The state of the art for this type of filters is XOR [\cite{graf2020xor}] and Fuse filters [\cite{graf2022binary}], with yet no many contributions dedicated to \kmers, e.g. this exception~[\cite{ulrich2024fast}] in taxonomic assignment.
\subsubsection{Other filters}
\paragraph{\textbf{Quotient filters}} are another structure used for approximate set membership tests, similar to Bloom filters. It provides efficient insert, delete, and membership query operations with controlled false positive rates and space efficiency. Quotient filters leverage quotienting (see Figure \ref{fig:amq} right):  
the hash value is split in two parts, high order (prefix, or quotient) and low order (suffix, or remainder) bits. Quotient bits address \kmers in the structure, while remainders are stored as fingerprints. The filter uses an array of buckets, where each bucket can store multiple entries. Quotient filters use less space than traditional hash tables and can be more space-efficient than Bloom filters in certain scenarios. They support dynamic operations (insertion and deletion) more seamlessly than Bloom filters. 
Recent advances in quotient filters mitigate performance degradation as they fill up~[\cite{pandey2021vector}], enhancing large scale use, or specialize in associating counts to \kmers~[\cite{pandey2017general, pandey2018squeakr, levallois2024backpack}] and associated to a de Bruijn graph [\cite{Pandey2017}]. 

\paragraph{\textbf{Cuckoo filters}} stores fingerprints of elements in a table with multiple possible locations, which reduces the risk of collisions. If a collision occurs, the element can "kick out" (as cuckoos do) an existing element to another location, making cuckoo filters more space-efficient than Bloom filters. In \cite{Zentgraf2020}, an advanced type of cuckoo filters is chosen because it maintains cache efficiency when dealing with gapped \kmers, which are \kmers separated by a fixed number of nucleotides (gaps) rather than being consecutive.\\


\paragraph{\textbf{Example use cases of filters}}
Bloom filters are spread in a very large number of bioinformatics tools, often used for \kmer pre-filtering. Interestingly, some tools can fully handle the false positives they produce and end up being exact, such as the assembler Minia that detects the false positives thanks to the de Bruijn graph structure, or khmer [\cite{crusoe2015khmer}]. Bifrost [\cite{holley2019bifrost}] builds unitigs and uses Bloom filters for speed-up and corrects a posteriori erroneous \kmers. Aside from assembly, Minia was used in [\cite{krannich2022population}], a study involves analyzing genomic sequences from diverse human populations to detect non-reference variants absent from the human reference GRCh38.\\
A study on Ossabaw minipigs, that are genetically predisposed to a metabolic syndrome that increase the risk of heart disease, stroke, and diabetes, used a \kmer approach based on cuckoo filters that identified differences in clusters of genes encoding mitochondrial and inflammatory proteins [\cite{kleinbongard2022non}].

\subsection{Hash tables}\label{section:hashtables}
\subsubsection{Dynamic hash tables}

Hash tables are dictionaries associating keys and values (e.g., genomes of origin) using hash functions. Dictionaries typically store their key sets and the associated values. In our case, a \kmer is hashed using a hash function, which provides an address in the hash table where it handled to be stored as a key with its associated value.

General purpose hash tables are used for \kmers whenever dynamicity is required and the cost in bit per \kmer is not a bottleneck. Among very competitive options, Rust's HashSet is based on state-of-the-art Swiss Tables, and other possibilities and their different trade-offs benefit from a comprehensive benchmark\footnote{\url{https://github.com/martinus/map_benchmark}}. 
\cite{diaz2024space} present a space-efficient method for counting \kmers that involves a \kmer hash table. It presents similarity with SPSS and also the SBWT approach, as it uses k-1 overlaps between consecutive \kmers and records only some of the de Bruijn graph edges in its structure. Almost each entry \kmer only records its last symbol and a pointer to the previous \kmer, which allows to quickly retrieve and update counts for repeated patterns in the data.

\subsubsection{Static hash tables}

\begin{figure}[ht]
    \centering
    \includegraphics[width=0.8\textwidth]{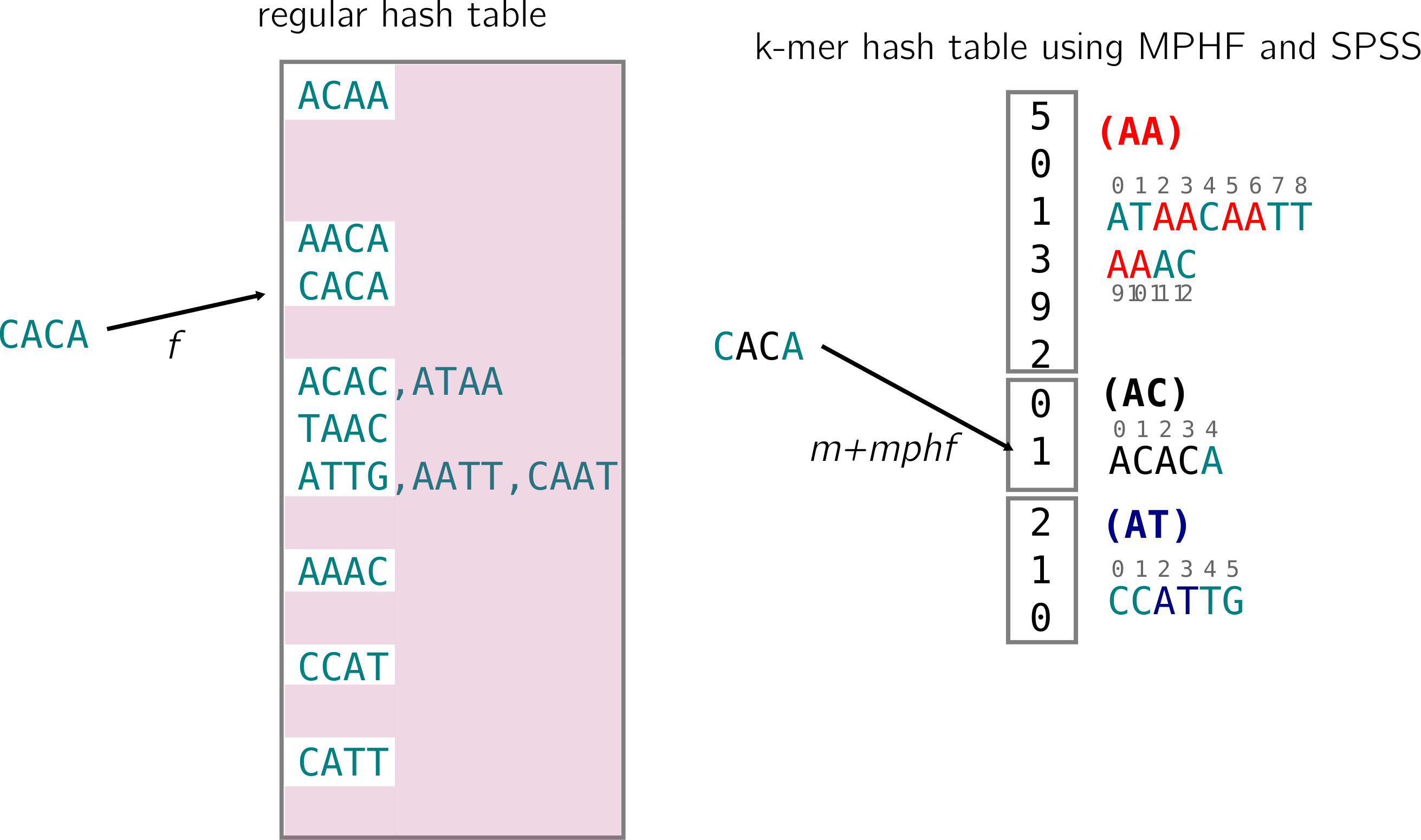}
    \caption{Comparison of a regular hash table and a \kmer hash table using a MPHF. In the regular hash table on the left, bits allocated for the red part participate in a larger overhead than for MPHF-based methods. They ensure the possibility to add new elements while keeping an efficient query, and manage collisions. Keys are stored independently. On the left, a \kmer is addressed to a bucket thanks to its minimizer (here lexicographic, of size 2), and to store the key, a MPHF associates the \kmer to its position in a SPSS, so several keys are co-encoded.}
    \label{fig:hash_table}
\end{figure}

\paragraph{\textbf{Minimal perfect hash functions}} Efficient hash functions (minimal perfect hash functions, MPHF) have been developed to associate keys with values. Contrary to regular hash functions MPHFs allocate space exactly for the required number of distinct \kmers. Regular hash tables, on the contrary, are dynamic and allocate space according to a desired load factor (Figure \ref{fig:hash_table}).
However MPHFs cannot handle alien keys once built for a given set, which means that MPHFS coupled with a system to store exact keys can lead to hash tables, while MPHFs coupled with lossy fingerprints have a filter behavior (see for instance [\cite{yu2018seqothello}]).

When MPHFs help create hash tables, such as BLight~[\cite{marchet2021blight}] or SSHash~[\cite{pibiri2022sparse}], they are capable of \kmer presence/absence queries and associating additional information with \kmers. 

In these hash tables, the second key ingredient is the technique to store the key set. 
One drawback of the MPHFs is that they must be fed a set of distinct \kmers, they cannot handle a regular FASTA.
In order to leverage the redundancy in thes \kmer set to be stored and to provide a\kmer set, SPSS can be used. They have the property to store keys compactly. One of the first examples leveraging this principle is the Pufferfish \kmer index \cite{almodaresi2018space}, using unitigs and a MPHF. Most efficient hash tables based on MPHF achieve storing 31-mers of a human genome in around 2-3 GB.

\begin{figure}[ht]
    \centering
    \includegraphics[width=0.7\textwidth]{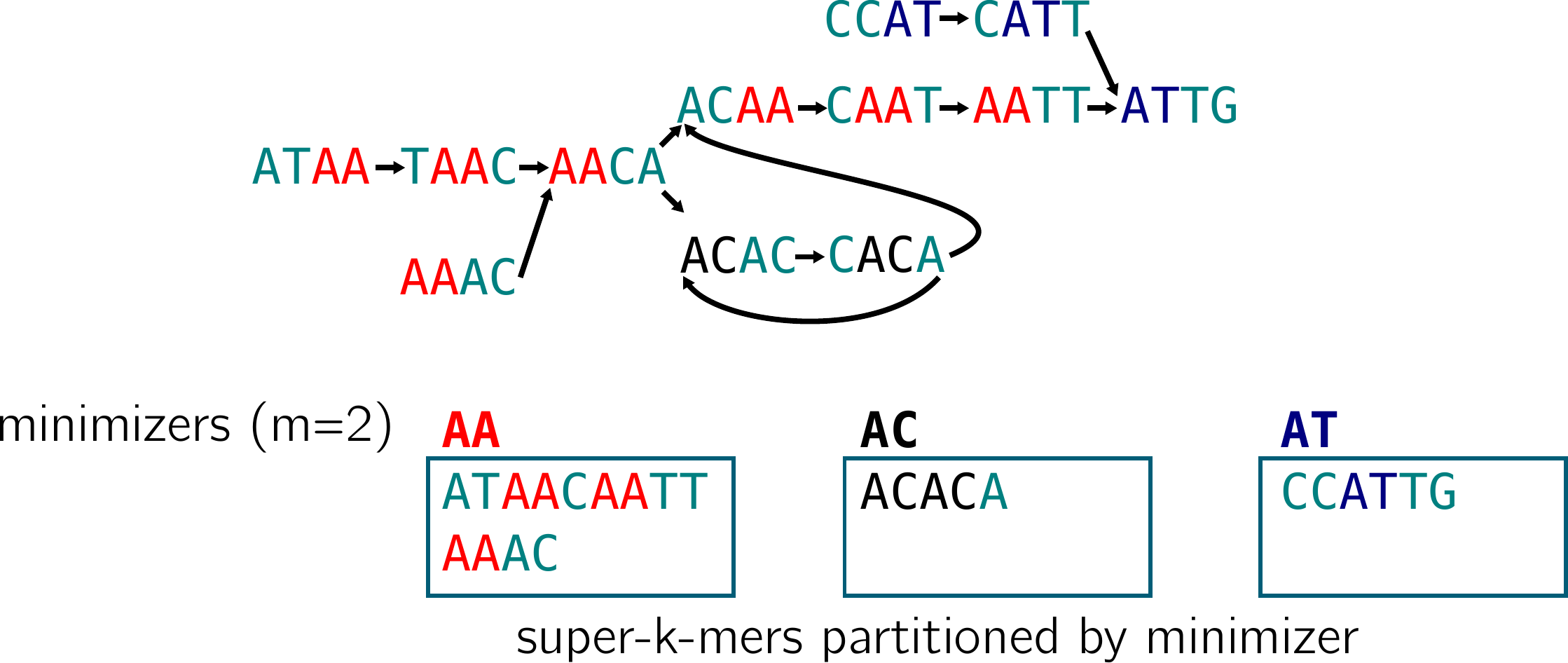}
    \caption{Example of super-\kmers built with a lexicographical minimizer of size 2.}
    \label{fig:spkm}
\end{figure}

\paragraph{\textbf{Minimizer partitioning}} Minimizers~[\cite{roberts2004reducing,schleimer2003winnowing}] are the product of selecting the smallest m-mer within a \kmer. Several interesting properties ensue, such as the fact that consecutive \kmers are likely to share a minimizer. Practically, minimizers are not selected on a lexicographic order, but by hashing m-mers and selecting the smallest integer. This has empirically proven to have better distribution properties [\cite{chikhi2014representation}] for the minimizers. Minimizer induce a natural and deterministic partition of the \kmers, by grouping \kmers in $4^{m}$ buckets according to their minimizer. This property is helpful to reduce encoding integer sizes and parallelism. They are used coupled to MPHFs for instance in [\cite{bcalm2,pibiri2022sparse}].\\

\paragraph{\textbf{Super-\emph{k}-mers}} Tools like BLight utilize a special form of SPSS, the super-\kmers. These super-\kmers group consecutive \kmers and naturally partition \kmer sets, facilitating smaller integer usage and parallelization. 
They are an interesting case of SPSS, as they are driven by hashing techniques: 
super-\kmers are based on hashed minimizers (Figure \ref{fig:spkm}).\\
 Since consecutive \kmers stay close in the structure, this has positive impact on the query speed, as groups of \kmers in the query are queried almost simultaneously. It is worth mentioning that super-$k$-mers can be built from reads, unitigs, or other longer strings. Other properties for compression emerge when several samples are stored and compressed together using a colored de Bruijn graph, because consecutive \kmers are likely to share similar information [\cite{Pibiri2024,karasikov2020metagraph}].\\

\paragraph{\textbf{Example use cases of hash tables}}
Hash tables are also a widespread data structure across bioinformatics. They are a method of choice for many unitig builders, such as BCALM2 [\cite{bcalm2}] or Cuttlefish2 [\cite{khan2022scalable}]. They are oftentimes embedded in colored de Bruijn graphs for joint \kmer analysis of multiple samples such as in [\cite{fan2024fulgor, marchet2020reindeer}], but also in alignment-free methods [\cite{almodaresi2021puffaligner}], and alignment free methods for RNA-seq quantification [\cite{patro2017salmon}] (using a hash table on the cuckoo principle). 

The minimizer partitioning technique spread across sequence bioinformatics, it is for instance used in \kmer based metagenomic classifier Kraken2~[\cite{wood2019improved}], Kmtricks ~\cite{10.1093/bioadv/vbac029} or in the Kmer File Format (KFF) \kmer manager [\cite{dufresne2022k}]. Associated to super-\kmers built from reads, it is also behind the partitioning of the KMC2\cite{deorowicz2015kmc} \kmer counter, that uses a disk-based sort-and-merge paradigm.


\section{Conclusion}
\subsection{Summary}

The impressive advancements in sequencing technologies are often rightfully praised, including in popular science. However, \kmer-based methods, which are crucial for effectively handling this data, are frequently dismissed as mere technical details. These methods, however, represent significant achievements, combining advanced algorithms and excellent software engineering. They enable us to store the entire human genome's sequence in just a few gigabytes of RAM—a capacity that can fit into a modern smartphone.

\begin{figure}[ht]
    \centering
    \includegraphics[width=1.1\textwidth]{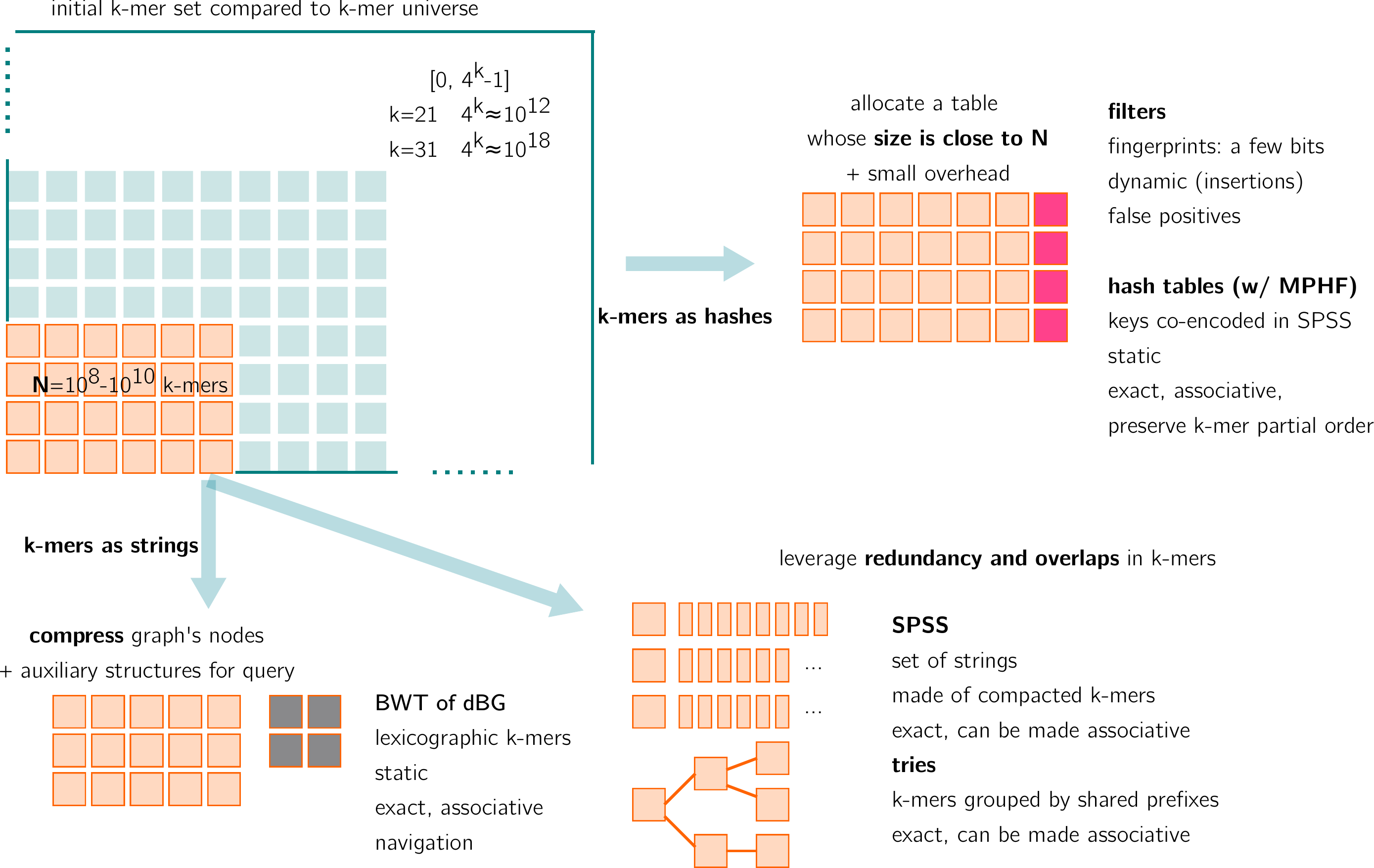}
    \caption{Summary of methodological choices in \kmer sets representations}
    \label{fig:summary_kmers}
\end{figure}

In the current state of the art, I identify two main strategies for \kmer set representation (Figure \ref{fig:summary_kmers}): 
\begin{enumerate}
    \item[$\diamond$] Storing \kmer fingerprints in slots, with space allocated closely matching the real number of \kmers, using small fingerprints with potential false positives (filters) or co-encoded (hash tables based on SPSS).
    \item[$\diamond$] Using lexicographic information of the \kmers or the graph, with transforms or algorithms that re-arranges the \kmer information in order to reduce the size of the representation.
    \item[$\diamond$] These two strategies, or sub-strategies, can be mixed. Typically, entry \kmers are treated as hashes in hash tables but efficient hash tables store their keys in compact way using lexicographic properties. Another example is is the super-\kmers, that leverage hash functions to create \kmer superstrings (compact lexicographic representations) for efficient \kmer partitioning.
\end{enumerate}

Ten years back, mostly probabilistic methods could allow going below the information-theory space lower bound, with a few exact methods, at the price of offering a shallow range of operations. Nowadays, when solely looking at the bit per \kmer footprint, recent most well-performing methods are neck and neck and below the lower bound, and provide membership and associativity. Single queries are in the order of hundreds of nanoseconds.

Key differences include that SPSS-based solutions preserve partial \kmer order, an important property for quick queries in hash tables and compression in colored structures. BWT-based methods provide associative, highly compressed structures yielding \kmer ranks. Filters and probabilistic methods are still being used because they allow some dynamicity (\kmer insertion) and are especially quick to build. However, recent advances offer fully dynamic and exact solutions using slightly more space.

\begin{figure}[ht]
    \centering
    \includegraphics[width=\textwidth]{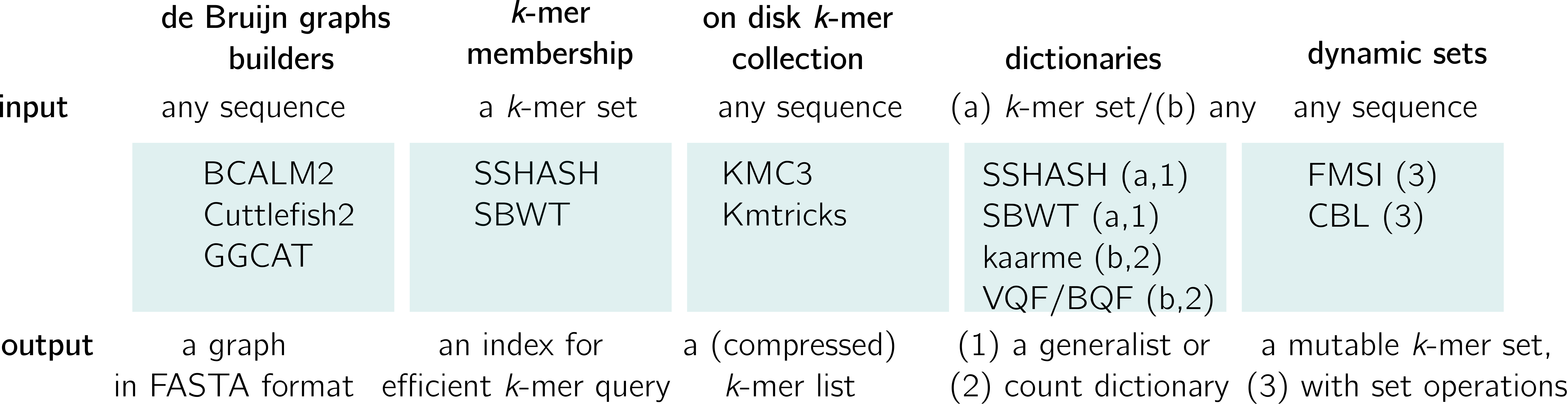}
    \caption{Some \kmer structures seen through their functional aspects. Among tools that specialize in building de Bruijn graph sequences, BCALM2 focuses on being memory lightweight, 
Cuttelfish2 on scaling to large size of inputs, and GGCAT  on speed.
BCALM2 and Cuttelfish output a unitig graph, GGCAT has several SPSS options. BCALM2 and Cuttlefish yield a graph written on disk, GGCAT allows to load it in RAM for queries. Cuttlefish (and also TwoPaCo, have an option to speed-up construction if sequences are an assemble genome). $K$-mer indexes, dictionaries and sets can be loaded in RAM for membership queries.
In \kmer indexes, SSHash is memory efficient and focuses on query speed, and the bit-matrix SBWT provides compression. }
    \label{fig:practical}
\end{figure}

About practical tool usage, a confusing aspect is that many methods call themselves de Bruijn graphs but do not actually output sequences of a de Bruijn graph. This is because they rather consider the de Bruijn graph as an inner representation for the \kmer sets than about a final product. I give a summary in Figure \ref{fig:practical}. Not mentioned in details, some \kmer tools act as \kmer managers (writing \kmers on disk, partitioning \kmers \ldots) such as [\cite{crusoe2015khmer,dufresne2022k}].

\subsection{Trends and future directions for \emph{k}-mer sets}

\noindent\textbf{Are we hitting a wall in terms of efficiency?} We don't have many informative lower bounds, which makes it difficult to answer. The community is still exploring, notably by discovering new structural properties of \kmers and sets~[\cite{Abrar2024.02.08.579510}]. Currently, mostly lexicographic transforms based on rotations are used, but other approaches could be tested for better representation. Another option, learned indexes~[\cite{ferragina2020pgm}], is being explored to capture structural properties for optimized representation and queries. They can also be used for compression, especially for unassembled data for which there is room for improvement.
Then, other techniques from neighbour computer science fields can inspire techniques close to minimal perfect hash functions (MPHFs) and Bloom filters, that could be expanded to other techniques close to filters and MPHF (such as XOR and Fuse filters, Othello~[\cite{graf2020xor,graf2022binary,yu2018memory}]). In order to compare all those techniques and the future ones, a comprehensive benchmark comparing performance in terms of construction time, query time, and disk/RAM usage, based on the size of the initial set and the value of $k$, is still lacking.
\paragraph{\textbf{Good \emph{k}-mer partitions.}} More technical but critical for performances, the improvement of the computation of minimizers and super-\kmers is also actively researched~[\cite{kunzmann2024fast,Koerkamp2024.05.25.595898,hoang2024density}]. Recently, it lead to proposing the first minimal perfect hash function dedicated to \kmers [\cite{pibiri2023locality}], that can go below the space lower bound of general-purpose MPHF in certain scenarios. One interesting direction is $k$-mer partitioning using minimizers. Assigning a k-mer to its minimizer is a natural partition, but in practice, it is not very uniform in the target space, even when using hashing. This results in additional resource costs for large datasets. The solutions found are generally relegated to appendices and are minimally or not evaluated. We still lack a more theoretical framework on this issue.
\paragraph{\textbf{Query driven developments}} Using hashing is essential to fill structures that allocate tables or bit arrays. However, hashing loses the lexicographic information that \kmers carry. Not only two consecutive \kmers are typically parted, but it makes queries less informative in terms of distance between the sequence and the content of the index. Several hybrids (hash+string representation of \kmers, trie+quotienting, \ldots) try to mitigate the problem, but some are to my knowledge not yet explored, such as Bloom filters + SPSS. Another promising approach is to record partial lexicographic information in fingerprints that have properties close to hash [\cite{greenberg2023lexichash,Shen2024.08.30.610459}]. It is especially interesting coupled to applications where \kmers are sampled and very sparse, where string-based approaches are less helpful.
\paragraph{\textbf{Longer \emph{k}-mers and different alphabets}} The field is partially shaped by the type of sequencing data that becomes dominantly used and indexed. The growing throughput of sequencing data also motivates the possibility to index \kmers in streaming, therefore operations are expanding to include insertions and set operations. Long reads also become more and more accurate, which should motivate algorithmic solutions for very large \kmers in the future, as well as extended alphabet supporting IUPAC alphabets, since more and more modified bases are being called in reads such as from Oxford Nanopore.
\paragraph{\textbf{\emph{K}-mer based tools for bioinformatics analysis}}
Finally, in a companion paper [\cite{Marchet2024colored}], I review one direct application for \kmer set structures: structures than aggregate different sets from multiple datasets or genomes, called colored \kmer sets.
\section*{Fundings}

This study has been supported by ANR JCJC Find-RNA [ANR -23-CE45-0003-01].

\section*{Conflict of interest disclosure}

The author declares that she complies with the PCI rule of having no financial conflicts of interest in relation to the content of the article.

\section*{Data, script, code, and supplementary information availability}
None declared.

\section*{Acknowledgments}
This manuscript and its companion were created using my notes from talks I gave in recent conferences, courses and workshops. I'd like to thank the community for inviting me and giving me a opportunity to present and discuss my views on those subjects.  I'd also like to thanks J. Alanko, G. Pibiri, A. Limasset and L. Robidou for going over the manuscript.

\bibliography{arxiv}

\begin{thebibliography}{90}
\providecommand{\natexlab}[1]{#1}
\providecommand{\url}[1]{\texttt{#1}}
\expandafter\ifx\csname urlstyle\endcsname\relax
  \providecommand{\doi}[1]{doi: #1}\else
  \providecommand{\doi}{doi: \begingroup \urlstyle{rm}\Url}\fi

\bibitem[Abrar and Medvedev(2024)]{Abrar2024.02.08.579510}
Md.~Hasin Abrar and Paul Medvedev.
\newblock Pla-complexity of k-mer multisets.
\newblock \emph{bioRxiv}, 2024.
\newblock \doi{10.1101/2024.02.08.579510}.
\newblock URL
  \url{https://www.biorxiv.org/content/early/2024/02/11/2024.02.08.579510}.

\bibitem[Agret et~al.(2022)Agret, Chateau, Droc, Sarah, Ruiz, and
  Mancheron]{agret2020redoak}
Clément Agret, Annie Chateau, Gaetan Droc, Gautier Sarah, Manuel Ruiz, and
  Alban Mancheron.
\newblock Redoak: a reference-free and alignment-free structure for indexing a
  collection of similar genomes.
\newblock \emph{Journal of Open Source Software}, 7\penalty0 (80):\penalty0
  4363, 2022.
\newblock \doi{10.21105/joss.04363}.
\newblock URL \url{https://doi.org/10.21105/joss.04363}.

\bibitem[Alanko et~al.(2021)Alanko, Alipanahi, Settle, Boucher, and
  Gagie]{alanko2021buffering}
Jarno Alanko, Bahar Alipanahi, Jonathen Settle, Christina Boucher, and Travis
  Gagie.
\newblock Buffering updates enables efficient dynamic de bruijn graphs.
\newblock \emph{Computational and structural biotechnology journal},
  19:\penalty0 4067--4078, 2021.

\bibitem[Alanko et~al.(2023{\natexlab{a}})Alanko, Biagi, and
  Puglisi]{alanko2023longest}
Jarno~N Alanko, Elena Biagi, and Simon~J Puglisi.
\newblock Longest common prefix arrays for succinct k-spectra.
\newblock In \emph{International Symposium on String Processing and Information
  Retrieval}, pages 1--13. Springer, 2023{\natexlab{a}}.

\bibitem[Alanko et~al.(2023{\natexlab{b}})Alanko, Puglisi, and
  Vuohtoniemi]{alanko2023small}
Jarno~N Alanko, Simon~J Puglisi, and Jaakko Vuohtoniemi.
\newblock Small searchable $\kappa$-spectra via subset rank queries on the
  spectral burrows-wheeler transform.
\newblock In \emph{SIAM Conference on Applied and Computational Discrete
  Algorithms (ACDA23)}, pages 225--236. SIAM, 2023{\natexlab{b}}.

\bibitem[Alipanahi et~al.(2020)Alipanahi, Muggli, Jundi, Noyes, and
  Boucher]{alipanahi2020metagenome}
Bahar Alipanahi, Martin~D Muggli, Musa Jundi, Noelle~R Noyes, and Christina
  Boucher.
\newblock Metagenome snp calling via read-colored de bruijn graphs.
\newblock \emph{Bioinformatics}, 36\penalty0 (22-23):\penalty0 5275--5281,
  2020.

\bibitem[Alipanahi et~al.(2021)Alipanahi, Kuhnle, Puglisi, Salmela, and
  Boucher]{alipanahi2021succinct}
Bahar Alipanahi, Alan Kuhnle, Simon~J Puglisi, Leena Salmela, and Christina
  Boucher.
\newblock Succinct dynamic de bruijn graphs.
\newblock \emph{Bioinformatics}, 37\penalty0 (14):\penalty0 1946--1952, 2021.

\bibitem[Almodaresi et~al.(2018)Almodaresi, Sarkar, Srivastava, and
  Patro]{almodaresi2018space}
Fatemeh Almodaresi, Hirak Sarkar, Avi Srivastava, and Rob Patro.
\newblock A space and time-efficient index for the compacted colored de bruijn
  graph.
\newblock \emph{Bioinformatics}, 34\penalty0 (13):\penalty0 i169--i177, 2018.

\bibitem[Almodaresi et~al.(2021)Almodaresi, Zakeri, and
  Patro]{almodaresi2021puffaligner}
Fatemeh Almodaresi, Mohsen Zakeri, and Rob Patro.
\newblock Puffaligner: a fast, efficient and accurate aligner based on the
  pufferfish index.
\newblock \emph{Bioinformatics}, 37\penalty0 (22):\penalty0 4048--4055, 2021.

\bibitem[Bankevich et~al.(2012)Bankevich, Nurk, Antipov, Gurevich, Dvorkin,
  Kulikov, Lesin, Nikolenko, Pham, Prjibelski, et~al.]{bankevich2012spades}
Anton Bankevich, Sergey Nurk, Dmitry Antipov, Alexey~A Gurevich, Mikhail
  Dvorkin, Alexander~S Kulikov, Valery~M Lesin, Sergey~I Nikolenko, Son Pham,
  Andrey~D Prjibelski, et~al.
\newblock Spades: a new genome assembly algorithm and its applications to
  single-cell sequencing.
\newblock \emph{Journal of computational biology}, 19\penalty0 (5):\penalty0
  455--477, 2012.

\bibitem[Bonin et~al.(2023)Bonin, Doster, Worley, Pinnell, Bravo, Ferm, Marini,
  Prosperi, Noyes, Morley, et~al.]{bonin2023megares}
Nathalie Bonin, Enrique Doster, Hannah Worley, Lee~J Pinnell, Jonathan~E Bravo,
  Peter Ferm, Simone Marini, Mattia Prosperi, Noelle Noyes, Paul~S Morley,
  et~al.
\newblock Megares and amr++, v3. 0: an updated comprehensive database of
  antimicrobial resistance determinants and an improved software pipeline for
  classification using high-throughput sequencing.
\newblock \emph{Nucleic acids research}, 51\penalty0 (D1):\penalty0 D744--D752,
  2023.

\bibitem[Bowe et~al.(2012)Bowe, Onodera, Sadakane, and
  Shibuya]{bowe2012succinct}
Alexander Bowe, Taku Onodera, Kunihiko Sadakane, and Tetsuo Shibuya.
\newblock Succinct de bruijn graphs.
\newblock In Ben Raphael and Jijun Tang, editors, \emph{Algorithms in
  Bioinformatics}, pages 225--235, Berlin, Heidelberg, 2012. Springer Berlin
  Heidelberg.
\newblock ISBN 978-3-642-33122-0.

\bibitem[Brinda(2016)]{brinda2016nouvelles}
Karel Brinda.
\newblock \emph{Nouvelles techniques informatiques pour la localisation et la
  classification de donn{\'e}es de s{\'e}quen{\c{c}}age haut d{\'e}bit}.
\newblock PhD thesis, Paris Est, 2016.

\bibitem[B{\v{r}}inda et~al.(2021)B{\v{r}}inda, Baym, and
  Kucherov]{bvrinda2021simplitigs}
Karel B{\v{r}}inda, Michael Baym, and Gregory Kucherov.
\newblock Simplitigs as an efficient and scalable representation of de bruijn
  graphs.
\newblock \emph{Genome biology}, 22:\penalty0 1--24, 2021.

\bibitem[Bushmanova et~al.(2019)Bushmanova, Antipov, Lapidus, and
  Prjibelski]{bushmanova2019rnaspades}
Elena Bushmanova, Dmitry Antipov, Alla Lapidus, and Andrey~D Prjibelski.
\newblock rnaspades: a de novo transcriptome assembler and its application to
  rna-seq data.
\newblock \emph{GigaScience}, 8\penalty0 (9):\penalty0 giz100, 2019.

\bibitem[Chikhi(2021)]{chikhi2021tale}
Rayan Chikhi.
\newblock A tale of optimizing the space taken by de bruijn graphs.
\newblock In \emph{Connecting with Computability: 17th Conference on
  Computability in Europe, CiE 2021, Virtual Event, Ghent, July 5--9, 2021,
  Proceedings 17}, pages 120--134. Springer, 2021.

\bibitem[Chikhi et~al.(2014)Chikhi, Limasset, Jackman, Simpson, and
  Medvedev]{chikhi2014representation}
Rayan Chikhi, Antoine Limasset, Shaun Jackman, Jared~T. Simpson, and Paul
  Medvedev.
\newblock On the representation of de bruijn graphs.
\newblock In Roded Sharan, editor, \emph{Research in Computational Molecular
  Biology}, pages 35--55, Cham, 2014. Springer International Publishing.
\newblock ISBN 978-3-319-05269-4.

\bibitem[Chikhi et~al.(2016)Chikhi, Limasset, and Medvedev]{bcalm2}
Rayan Chikhi, Antoine Limasset, and Paul Medvedev.
\newblock {Compacting de Bruijn graphs from sequencing data quickly and in low
  memory}.
\newblock \emph{Bioinformatics}, 32\penalty0 (12):\penalty0 i201--i208, 06
  2016.
\newblock ISSN 1367-4803.
\newblock \doi{10.1093/bioinformatics/btw279}.
\newblock URL \url{https://doi.org/10.1093/bioinformatics/btw279}.

\bibitem[Chikhi et~al.(2019)Chikhi, Holub, and Medvedev]{chikhi2021data}
Rayan Chikhi, Jan Holub, and Paul Medvedev.
\newblock Data structures to represent a set of k-long dna sequences.
\newblock \emph{ACM Computing Surveys}, 54\penalty0 (1):\penalty0 17:1--17:22,
  2019.

\bibitem[Conway and Bromage(2011)]{conway2011succinct}
Thomas~C Conway and Andrew~J Bromage.
\newblock Succinct data structures for assembling large genomes.
\newblock \emph{Bioinformatics}, 27\penalty0 (4):\penalty0 479--486, 2011.

\bibitem[Cracco and Tomescu(2023)]{cracco2023extremely}
Andrea Cracco and Alexandru~I Tomescu.
\newblock Extremely fast construction and querying of compacted and colored de
  bruijn graphs with ggcat.
\newblock \emph{Genome Research}, pages gr--277615, 2023.

\bibitem[Crusoe et~al.(2015)Crusoe, Alameldin, Awad, Boucher, Caldwell,
  Cartwright, Charbonneau, Constantinides, Edvenson, Fay,
  et~al.]{crusoe2015khmer}
Michael~R Crusoe, Hussien~F Alameldin, Sherine Awad, Elmar Boucher, Adam
  Caldwell, Reed Cartwright, Amanda Charbonneau, Bede Constantinides, Greg
  Edvenson, Scott Fay, et~al.
\newblock The khmer software package: enabling efficient nucleotide sequence
  analysis.
\newblock \emph{F1000Research}, 4, 2015.

\bibitem[Dadi et~al.(2018)Dadi, Siragusa, Piro, Andrusch, Seiler, Renard, and
  Reinert]{dadi2018dream}
Temesgen~Hailemariam Dadi, Enrico Siragusa, Vitor~C Piro, Andreas Andrusch,
  Enrico Seiler, Bernhard~Y Renard, and Knut Reinert.
\newblock Dream-yara: an exact read mapper for very large databases with short
  update time.
\newblock \emph{Bioinformatics}, 34\penalty0 (17):\penalty0 i766--i772, 2018.

\bibitem[Deorowicz et~al.(2015)Deorowicz, Kokot, Grabowski, and
  Debudaj-Grabysz]{deorowicz2015kmc}
Sebastian Deorowicz, Marek Kokot, Szymon Grabowski, and Agnieszka
  Debudaj-Grabysz.
\newblock Kmc 2: fast and resource-frugal k-mer counting.
\newblock \emph{Bioinformatics}, 31\penalty0 (10):\penalty0 1569--1576, 2015.

\bibitem[D{\'i}az-Dom{\'i}nguez et~al.(2024)D{\'i}az-Dom{\'i}nguez, Leinonen,
  and Salmela]{diaz2024space}
Diego D{\'i}az-Dom{\'i}nguez, Miika Leinonen, and Leena Salmela.
\newblock Space-efficient computation of k-mer dictionaries for large values of
  k.
\newblock \emph{Algorithms for Molecular Biology}, 19\penalty0 (1):\penalty0
  14, 2024.

\bibitem[Dufresne et~al.(2022)Dufresne, Lemane, Marijon, Peterlongo, Rahman,
  Kokot, Medvedev, Deorowicz, and Chikhi]{dufresne2022k}
Yoann Dufresne, Teo Lemane, Pierre Marijon, Pierre Peterlongo, Amatur Rahman,
  Marek Kokot, Paul Medvedev, Sebastian Deorowicz, and Rayan Chikhi.
\newblock The k-mer file format: a standardized and compact disk representation
  of sets of k-mers.
\newblock \emph{Bioinformatics}, 38\penalty0 (18):\penalty0 4423--4425, 2022.

\bibitem[Fan et~al.(2024)Fan, Khan, Singh, et~al.]{fan2024fulgor}
J.~Fan, J.~Khan, N.P. Singh, et~al.
\newblock Fulgor: a fast and compact k-mer index for large-scale matching and
  color queries.
\newblock \emph{Algorithms for Molecular Biology}, 19:\penalty0 3, 2024.
\newblock \doi{10.1186/s13015-024-00251-9}.
\newblock URL \url{https://doi.org/10.1186/s13015-024-00251-9}.

\bibitem[Fan et~al.(2000)Fan, Cao, Almeida, and Broder]{Fan2000}
Li~Fan, Pei Cao, Jussara Almeida, and Andrei~Z. Broder.
\newblock Summary cache: a scalable wide-area web cache sharing protocol.
\newblock \emph{IEEE/ACM Transactions on Networking (TON)}, 8\penalty0
  (3):\penalty0 281--293, 2000.
\newblock \doi{10.1109/90.851975}.

\bibitem[Ferragina and Manzini(2005)]{ferragina2005fmindex}
Paolo Ferragina and Giovanni Manzini.
\newblock Indexing compressed text.
\newblock \emph{J. {ACM}}, 52\penalty0 (4):\penalty0 552--581, 2005.

\bibitem[Ferragina and Vinciguerra(2020)]{ferragina2020pgm}
Paolo Ferragina and Giorgio Vinciguerra.
\newblock The pgm-index: a fully-dynamic compressed learned index with provable
  worst-case bounds.
\newblock \emph{Proceedings of the VLDB Endowment}, 13\penalty0 (8):\penalty0
  1162--1175, 2020.

\bibitem[Graf and Lemire(2020)]{graf2020xor}
Thomas~Mueller Graf and Daniel Lemire.
\newblock Xor filters: Faster and smaller than bloom and cuckoo filters.
\newblock \emph{Journal of Experimental Algorithmics (JEA)}, 25:\penalty0
  1--16, 2020.

\bibitem[Graf and Lemire(2022)]{graf2022binary}
Thomas~Mueller Graf and Daniel Lemire.
\newblock Binary fuse filters: Fast and smaller than xor filters.
\newblock \emph{Journal of Experimental Algorithmics (JEA)}, 27\penalty0
  (1):\penalty0 1--15, 2022.

\bibitem[Greenberg et~al.(2023)Greenberg, Ravi, and
  Shomorony]{greenberg2023lexichash}
Grant Greenberg, Aditya~Narayan Ravi, and Ilan Shomorony.
\newblock Lexichash: sequence similarity estimation via lexicographic
  comparison of hashes.
\newblock \emph{Bioinformatics}, 39\penalty0 (11):\penalty0 btad652, 2023.

\bibitem[Hannoush et~al.(2024)Hannoush, Marchet, and
  Peterlongo]{hannoush2024cdbgtricks}
Khodor Hannoush, Camille Marchet, and Pierre Peterlongo.
\newblock Cdbgtricks: strategies to update a compacted de bruijn graph.
\newblock In \emph{Proceedings of the Prague Stringology Conference 2024 (PSC
  2024)}, pages 202--205. Czech Technical University in Prague, 2024.
\newblock URL \url{https://psc.fit.cvut.cz/event/2024/}.

\bibitem[Hoang et~al.(2024)Hoang, Marçais, and Kingsford]{hoang2024density}
M.~Hoang, G.~Marçais, and C.~Kingsford.
\newblock Density and conservation optimization of the generalized
  masked-minimizer sketching scheme.
\newblock \emph{Journal of Computational Biology: A Journal of Computational
  Molecular Cell Biology}, 31\penalty0 (1):\penalty0 2--20, 2024.
\newblock \doi{10.1089/cmb.2023.0212}.

\bibitem[Holley and Melsted(2019)]{holley2019bifrost}
Guillaume Holley and P{\'a}ll Melsted.
\newblock {Bifrost--Highly parallel construction and indexing of colored and
  compacted de Bruijn graphs}.
\newblock \emph{BioRxiv}, page 695338, 2019.

\bibitem[Holley et~al.(2016)Holley, Wittler, and Stoye]{holley2016bftrie}
Guillaume Holley, Roland Wittler, and Jens Stoye.
\newblock {Bloom Filter Trie}: an alignment-free and reference-free data
  structure for pan-genome storage.
\newblock \emph{Algorithms for Molecular Biology}, 11\penalty0 (1):\penalty0 3,
  2016.

\bibitem[Ingels et~al.(2024)Ingels, Martayan, Salson, and Marchet]{Ingels2024}
Florian Ingels, Igor Martayan, Mika\"el Salson, and Camille Marchet.
\newblock Constrained enumeration of k-mers from a collection of references
  with metadata.
\newblock \emph{bioRxiv}, 05 2024.
\newblock \doi{10.1101/2024.05.26.595967}.
\newblock URL \url{https://doi.org/10.1101/2024.05.26.595967}.

\bibitem[Iqbal et~al.(2012)Iqbal, Caccamo, Turner, Flicek, and
  McVean]{iqbal2012novo}
Zamin Iqbal, Mario Caccamo, Isaac Turner, Paul Flicek, and Gil McVean.
\newblock De novo assembly and genotyping of variants using colored de {Bruijn}
  graphs.
\newblock \emph{Nature genetics}, 44\penalty0 (2):\penalty0 226, 2012.

\bibitem[Jenike et~al.(2024)Jenike, Campos-Domínguez, Boddé, Cerca, Hodson,
  Schatz, and Jaron]{jenike2024guide}
Katharine~M. Jenike, Lucía Campos-Domínguez, Marilou Boddé, José Cerca,
  Christina~N. Hodson, Michael~C. Schatz, and Kamil~S. Jaron.
\newblock Guide to k-mer approaches for genomics across the tree of life.
\newblock \emph{arXiv preprint arXiv:2404.01519}, 2024.
\newblock \doi{10.48550/arXiv.2404.01519}.

\bibitem[Karasikov et~al.(2020)Karasikov, Mustafa, Danciu, Zimmermann, Barber,
  R{\"a}tsch, and Kahles]{karasikov2020metagraph}
Mikhail Karasikov, Harun Mustafa, Daniel Danciu, Marc Zimmermann, Christopher
  Barber, Gunnar R{\"a}tsch, and Andr{\'e} Kahles.
\newblock Metagraph: Indexing and analysing nucleotide archives at
  petabase-scale.
\newblock \emph{bioRxiv}, 2020.
\newblock \doi{10.1101/2020.10.01.322164}.
\newblock URL
  \url{https://www.biorxiv.org/content/early/2020/11/03/2020.10.01.322164}.

\bibitem[Khan et~al.(2022)Khan, Kokot, Deorowicz, and Patro]{khan2022scalable}
J.~Khan, M.~Kokot, S.~Deorowicz, and R.~Patro.
\newblock Scalable, ultra-fast, and low-memory construction of compacted de
  bruijn graphs with cuttlefish 2.
\newblock \emph{Genome Biology}, 23\penalty0 (1):\penalty0 190, 2022.
\newblock \doi{10.1186/s13059-022-02743-6}.

\bibitem[Kleinbongard et~al.(2022)Kleinbongard, Lieder, Skyschally, Alloosh,
  G{\"o}decke, Rahmann, Sturek, and Heusch]{kleinbongard2022non}
Petra Kleinbongard, Helmut~Raphael Lieder, Andreas Skyschally, Mouhamad
  Alloosh, Axel G{\"o}decke, Sven Rahmann, Michael Sturek, and Gerd Heusch.
\newblock Non-responsiveness to cardioprotection by ischaemic preconditioning
  in ossabaw minipigs with genetic predisposition to, but without the phenotype
  of the metabolic syndrome.
\newblock \emph{Basic research in cardiology}, 117\penalty0 (1):\penalty0 58,
  2022.

\bibitem[Koerkamp and Pibiri(2024)]{Koerkamp2024.05.25.595898}
Ragnar~Groot Koerkamp and Giulio~Ermanno Pibiri.
\newblock The mod-minimizer: a simple and efficient sampling algorithm for long
  k-mers.
\newblock \emph{bioRxiv}, 2024.
\newblock \doi{10.1101/2024.05.25.595898}.
\newblock URL
  \url{https://www.biorxiv.org/content/early/2024/07/07/2024.05.25.595898}.

\bibitem[Kokot et~al.(2017)Kokot, D{\l}ugosz, and Deorowicz]{kokot2017kmc}
Marek Kokot, Maciej D{\l}ugosz, and Sebastian Deorowicz.
\newblock Kmc 3: counting and manipulating k-mer statistics.
\newblock \emph{Bioinformatics}, 33\penalty0 (17):\penalty0 2759--2761, 2017.

\bibitem[Krannich et~al.(2022)Krannich, White, Niehus, Holley, Halld{\'o}rsson,
  and Kehr]{krannich2022population}
Thomas Krannich, W~Timothy~J White, Sebastian Niehus, Guillaume Holley,
  Bjarni~V Halld{\'o}rsson, and Birte Kehr.
\newblock Population-scale detection of non-reference sequence variants using
  colored de bruijn graphs.
\newblock \emph{Bioinformatics}, 38\penalty0 (3):\penalty0 604--611, 2022.

\bibitem[Kunzmann(2024)]{kunzmann2024fast}
Patrick Kunzmann.
\newblock A fast and simple approach to k-mer decomposition.
\newblock \emph{bioRxiv}, pages 2024--07, 2024.

\bibitem[Lemane et~al.(2022)Lemane, Medvedev, Chikhi, and
  Peterlongo]{10.1093/bioadv/vbac029}
Téo Lemane, Paul Medvedev, Rayan Chikhi, and Pierre Peterlongo.
\newblock {kmtricks: efficient and flexible construction of Bloom filters for
  large sequencing data collections}.
\newblock \emph{Bioinformatics Advances}, 2\penalty0 (1):\penalty0 vbac029, 04
  2022.
\newblock ISSN 2635-0041.
\newblock \doi{10.1093/bioadv/vbac029}.
\newblock URL \url{https://doi.org/10.1093/bioadv/vbac029}.

\bibitem[Levallois et~al.(2024)Levallois, Andreace, Le~Gal, Dufresne, and
  Peterlongo]{levallois2024backpack}
Victor Levallois, Francesco Andreace, Bertrand Le~Gal, Yoann Dufresne, and
  Pierre Peterlongo.
\newblock The backpack quotient filter: a dynamic and space-efficient data
  structure for querying k-mers with abundance.
\newblock \emph{bioRxiv}, pages 2024--02, 2024.

\bibitem[M\"aklin et~al.(2021)M\"aklin, Kallonen, Alanko, Samuelsen, Hegstad,
  M\"akinen, Corander, Heinz, and Honkela]{maklin2021bacterial}
Tommi M\"aklin, Teemu Kallonen, Jarno Alanko, {\O}rjan Samuelsen, Kristin
  Hegstad, Veli M\"akinen, Jukka Corander, Eva Heinz, and Antti Honkela.
\newblock Bacterial genomic epidemiology with mixed samples.
\newblock \emph{Microbial genomics}, 7\penalty0 (11):\penalty0 000691, 2021.

\bibitem[Marcais and Kingsford(2012)]{marcais2012jellyfish}
G~Marcais and C~Kingsford.
\newblock Jellyfish: A fast k-mer counter.
\newblock \emph{Tutorialis e Manuais}, 1\penalty0 (1-8):\penalty0 1038, 2012.

\bibitem[Marchet(2024)]{Marchet2024colored}
Camille Marchet.
\newblock Advancements in colored k-mer sets: essentials for the curious, 2024.
\newblock URL \url{https://arxiv.org/abs/2409.05214}.

\bibitem[Marchet et~al.(2020)Marchet, Iqbal, Gautheret, Salson, and
  Chikhi]{marchet2020reindeer}
Camille Marchet, Zamin Iqbal, Daniel Gautheret, Mika{\"e}l Salson, and Rayan
  Chikhi.
\newblock Reindeer: efficient indexing of k-mer presence and abundance in
  sequencing datasets.
\newblock \emph{Bioinformatics}, 36\penalty0 (Supplement\_1):\penalty0
  i177--i185, 2020.

\bibitem[Marchet et~al.(2021)Marchet, Kerbiriou, and
  Limasset]{marchet2021blight}
Camille Marchet, Mael Kerbiriou, and Antoine Limasset.
\newblock Blight: efficient exact associative structure for k-mers.
\newblock \emph{Bioinformatics}, 37\penalty0 (18):\penalty0 2858--2865, 2021.

\bibitem[Marini et~al.(2022)Marini, Mora, Boucher, Robertson~Noyes, and
  Prosperi]{marini2022towards}
Simone Marini, Rodrigo~A Mora, Christina Boucher, Noelle Robertson~Noyes, and
  Mattia Prosperi.
\newblock Towards routine employment of computational tools for antimicrobial
  resistance determination via high-throughput sequencing.
\newblock \emph{Briefings in bioinformatics}, 23\penalty0 (2):\penalty0
  bbac020, 2022.

\bibitem[Martayan et~al.(2024)Martayan, Cazaux, Limasset, and
  Marchet]{martayan2024conway}
Igor Martayan, Bastien Cazaux, Antoine Limasset, and Camille Marchet.
\newblock Conway-bromage-lyndon (cbl): an exact, dynamic representation of
  k-mer sets.
\newblock \emph{bioRxiv}, pages 2024--01, 2024.

\bibitem[Minkin et~al.(2016)Minkin, Pham, and
  Medvedev]{10.1093/bioinformatics/btw609}
Ilia Minkin, Son Pham, and Paul Medvedev.
\newblock {TwoPaCo: an efficient algorithm to build the compacted de Bruijn
  graph from many complete genomes}.
\newblock \emph{Bioinformatics}, 33\penalty0 (24):\penalty0 4024--4032, 09
  2016.
\newblock ISSN 1367-4803.
\newblock \doi{10.1093/bioinformatics/btw609}.
\newblock URL \url{https://doi.org/10.1093/bioinformatics/btw609}.

\bibitem[Nguyen et~al.(2021)Nguyen, Xue, Firlej, Ponty, Gallopin, and
  Gautheret]{nguyen2021reference}
Ha~TN Nguyen, Haoliang Xue, Virginie Firlej, Yann Ponty, Melina Gallopin, and
  Daniel Gautheret.
\newblock Reference-free transcriptome signatures for prostate cancer
  prognosis.
\newblock \emph{BMC cancer}, 21:\penalty0 1--12, 2021.

\bibitem[Pandey et~al.(2017{\natexlab{a}})Pandey, Bender, Johnson, and
  Patro]{pandey2017general}
Prashant Pandey, Michael~A Bender, Rob Johnson, and Rob Patro.
\newblock A general-purpose counting filter: Making every bit count.
\newblock In \emph{Proceedings of the 2017 ACM international conference on
  Management of Data}, pages 775--787, 2017{\natexlab{a}}.

\bibitem[Pandey et~al.(2017{\natexlab{b}})Pandey, Bender, Johnson, and
  Patwa]{Pandey2017}
Prashant Pandey, Michael~A. Bender, Rob Johnson, and Shikha Patwa.
\newblock debgr: an efficient and near-exact representation of the weighted de
  bruijn graph.
\newblock \emph{Bioinformatics}, 33\penalty0 (14):\penalty0 i133--i141,
  2017{\natexlab{b}}.
\newblock \doi{10.1093/bioinformatics/btx261}.
\newblock URL \url{https://doi.org/10.1093/bioinformatics/btx261}.

\bibitem[Pandey et~al.(2018)Pandey, Bender, Johnson, and
  Patro]{pandey2018squeakr}
Prashant Pandey, Michael~A Bender, Rob Johnson, and Rob Patro.
\newblock Squeakr: an exact and approximate k-mer counting system.
\newblock \emph{Bioinformatics}, 34\penalty0 (4):\penalty0 568--575, 2018.

\bibitem[Pandey et~al.(2021)Pandey, Conway, Durie, Bender, Farach-Colton, and
  Johnson]{pandey2021vector}
Prashant Pandey, Alex Conway, Joe Durie, Michael~A Bender, Martin
  Farach-Colton, and Rob Johnson.
\newblock Vector quotient filters: Overcoming the time/space trade-off in
  filter design.
\newblock In \emph{Proceedings of the 2021 International Conference on
  Management of Data}, pages 1386--1399, 2021.

\bibitem[Patro et~al.(2017{\natexlab{a}})Patro, Duggal, Love, Irizarry, and
  Kingsford]{patro2015salmon}
Rob Patro, Geet Duggal, Michael~I Love, Rafael~A Irizarry, and Carl Kingsford.
\newblock Salmon provides fast and bias-aware quantification of transcript
  expression.
\newblock \emph{Nature methods}, 14\penalty0 (4):\penalty0 417--419,
  2017{\natexlab{a}}.

\bibitem[Patro et~al.(2017{\natexlab{b}})Patro, Duggal, Love, Irizarry, and
  Kingsford]{patro2017salmon}
Rob Patro, Geet Duggal, Michael~I Love, Rafael~A Irizarry, and Carl Kingsford.
\newblock Salmon provides fast and bias-aware quantification of transcript
  expression.
\newblock \emph{Nature methods}, 14\penalty0 (4):\penalty0 417--419,
  2017{\natexlab{b}}.

\bibitem[Pibiri et~al.(2024)Pibiri, Fan, and Patro]{Pibiri2024}
Giulio~E. Pibiri, J.~Fan, and R.~Patro.
\newblock Meta-colored compacted de bruijn graphs.
\newblock In \emph{International Conference on Research in Computational
  Molecular Biology}, pages 131--146, Cham, 2024. Springer Nature Switzerland.

\bibitem[Pibiri(2022)]{pibiri2022sparse}
Giulio~Ermanno Pibiri.
\newblock Sparse and skew hashing of k-mers.
\newblock \emph{Bioinformatics}, 38\penalty0 (Supplement\_1):\penalty0
  i185--i194, 2022.

\bibitem[Pibiri et~al.(2023)Pibiri, Shibuya, and Limasset]{pibiri2023locality}
Giulio~Ermanno Pibiri, Yoshihiro Shibuya, and Antoine Limasset.
\newblock Locality-preserving minimal perfect hashing of k-mers.
\newblock \emph{Bioinformatics}, 39\penalty0 (Supplement\_1):\penalty0
  i534--i543, 2023.

\bibitem[Putze et~al.(2010)Putze, Sanders, and Singler]{putze2010cache}
Felix Putze, Peter Sanders, and Johannes Singler.
\newblock Cache-, hash-, and space-efficient bloom filters.
\newblock \emph{Journal of Experimental Algorithmics (JEA)}, 14:\penalty0 4--4,
  2010.

\bibitem[Rahman and Medevedev(2021)]{rahman2021representation}
Amatur Rahman and Paul Medevedev.
\newblock Representation of k-mer sets using spectrum-preserving string sets.
\newblock \emph{Journal of Computational Biology}, 28\penalty0 (4):\penalty0
  381--394, 2021.

\bibitem[Rahman and Medvedev(2022)]{rahman2022assembler}
Amatur Rahman and Paul Medvedev.
\newblock Assembler artifacts include misassembly because of unsafe unitigs and
  underassembly because of bidirected graphs.
\newblock \emph{Genome Research}, 32\penalty0 (9):\penalty0 1746--1753, 2022.

\bibitem[Riquier et~al.(2021)Riquier, Bessiere, Guibert, Bouge, Boureux,
  Ruffle, Audoux, Gilbert, Xue, Gautheret, et~al.]{riquier2021kmerator}
S{\'e}bastien Riquier, Chlo{\'e} Bessiere, Benoit Guibert, Anne-Laure Bouge,
  Anthony Boureux, Florence Ruffle, J{\'e}r{\^o}me Audoux, Nicolas Gilbert,
  Haoliang Xue, Daniel Gautheret, et~al.
\newblock Kmerator suite: design of specific k-mer signatures and automatic
  metadata discovery in large rna-seq datasets.
\newblock \emph{NAR genomics and bioinformatics}, 3\penalty0 (3):\penalty0
  lqab058, 2021.

\bibitem[Roberts et~al.(2004)Roberts, Hayes, Hunt, Mount, and
  Yorke]{roberts2004reducing}
Michael Roberts, Wayne Hayes, Brian~R Hunt, Stephen~M Mount, and James~A Yorke.
\newblock Reducing storage requirements for biological sequence comparison.
\newblock \emph{Bioinformatics}, 20\penalty0 (18):\penalty0 3363--3369, 2004.

\bibitem[Robidou and Peterlongo(2021)]{robidou2021findere}
Lucas Robidou and Pierre Peterlongo.
\newblock findere: fast and precise approximate membership query.
\newblock In \emph{String Processing and Information Retrieval: 28th
  International Symposium, SPIRE 2021, Lille, France, October 4--6, 2021,
  Proceedings 28}, pages 151--163. Springer, 2021.

\bibitem[Robidou and Peterlongo(2023)]{robidou2023fimpera}
Lucas Robidou and Pierre Peterlongo.
\newblock fimpera: drastic improvement of approximate membership query
  data-structures with counts.
\newblock \emph{Bioinformatics}, 39\penalty0 (5):\penalty0 btad305, 2023.

\bibitem[Rossignolo and Comin(2024)]{Rossignolo2024}
Enrico Rossignolo and Matteo Comin.
\newblock Enhanced compression of k-mer sets with counters via de bruijn
  graphs.
\newblock \emph{Journal of Computational Biology}, 2024.
\newblock \doi{to be updated}.
\newblock In press.

\bibitem[Sawada and Williams(2017)]{sawada2017practical}
Joe Sawada and Aaron Williams.
\newblock Practical algorithms to rank necklaces, lyndon words, and de bruijn
  sequences.
\newblock \emph{Journal of Discrete Algorithms}, 43:\penalty0 95--110, 2017.

\bibitem[Schleimer et~al.(2003)Schleimer, Wilkerson, and
  Aiken]{schleimer2003winnowing}
Saul Schleimer, Daniel~S Wilkerson, and Alex Aiken.
\newblock Winnowing: local algorithms for document fingerprinting.
\newblock In \emph{Proceedings of the 2003 ACM SIGMOD international conference
  on Management of data}, pages 76--85, 2003.

\bibitem[Schmidt and Alanko(2023)]{schmidt2023eulertigs}
Sebastian Schmidt and Jarno~N Alanko.
\newblock Eulertigs: minimum plain text representation of k-mer sets without
  repetitions in linear time.
\newblock \emph{Algorithms for Molecular Biology}, 18\penalty0 (1):\penalty0 5,
  2023.

\bibitem[Schmidt et~al.(2021)Schmidt, Khan, Alanko, and
  Tomescu]{schmidt2021matchtigs}
Sebastian Schmidt, Shahbaz Khan, Jarno Alanko, and Alexandru~I Tomescu.
\newblock Matchtigs: minimum plain text representation of kmer sets.
\newblock \emph{bioRxiv}, 2021.

\bibitem[Serajian et~al.(2024)Serajian, Marini, Alanko, Noyes, Prosperi, and
  Boucher]{10.1093/bioinformatics/btae243}
Mohammadali Serajian, Simone Marini, Jarno~N Alanko, Noelle~R Noyes, Mattia
  Prosperi, and Christina Boucher.
\newblock {Scalable de novo classification of antibiotic resistance of
  Mycobacterium tuberculosis}.
\newblock \emph{Bioinformatics}, 40\penalty0 (Supplement\_1):\penalty0
  i39--i47, 06 2024.
\newblock ISSN 1367-4811.
\newblock \doi{10.1093/bioinformatics/btae243}.
\newblock URL \url{https://doi.org/10.1093/bioinformatics/btae243}.

\bibitem[Shen and Iqbal(2024)]{Shen2024.08.30.610459}
Wei Shen and Zamin Iqbal.
\newblock Lexicmap: efficient sequence alignment against millions of
  prokaryotic genomes.
\newblock \emph{bioRxiv}, 2024.
\newblock \doi{10.1101/2024.08.30.610459}.
\newblock URL
  \url{https://www.biorxiv.org/content/early/2024/08/31/2024.08.30.610459}.

\bibitem[Shibuya et~al.(2022)Shibuya, Belazzougui, and
  Kucherov]{shibuya2022space}
Yoshihiro Shibuya, Djamal Belazzougui, and Gregory Kucherov.
\newblock Space-efficient representation of genomic k-mer count tables.
\newblock \emph{Algorithms for Molecular Biology}, 17\penalty0 (1):\penalty0 5,
  2022.

\bibitem[Sladk{\'y} et~al.(2023)Sladk{\'y}, Vesel{\'y}, and B{\v
  r}inda]{sladky2023masked}
Ond{\v r}ej Sladk{\'y}, Pavel Vesel{\'y}, and Karel B{\v r}inda.
\newblock Masked superstrings as a unified framework for textual k-mer set
  representations.
\newblock \emph{bioRxiv}, 2023.
\newblock \doi{10.1101/2023.02.01.526717}.
\newblock URL
  \url{https://www.biorxiv.org/content/early/2023/02/03/2023.02.01.526717}.

\bibitem[Sladk{\`y} et~al.(2024)Sladk{\`y}, Vesel{\`y}, and
  B{\v{r}}inda]{sladky2024function}
Ond{\v{r}}ej Sladk{\`y}, Pavel Vesel{\`y}, and Karel B{\v{r}}inda.
\newblock Function-assigned masked superstrings as a versatile and compact data
  type for k-mer sets.
\newblock \emph{bioRxiv}, pages 2024--03, 2024.

\bibitem[Ulrich and Renard(2024)]{ulrich2024fast}
Jens-Uwe Ulrich and Bernhard~Y Renard.
\newblock Fast and space-efficient taxonomic classification of long reads with
  hierarchical interleaved xor filters.
\newblock \emph{Genome Research}, pages gr--278623, 2024.

\bibitem[Wood and Salzberg(2014)]{wood2014kraken}
Derrick~E Wood and Steven~L Salzberg.
\newblock Kraken: ultrafast metagenomic sequence classification using exact
  alignments.
\newblock \emph{Genome biology}, 15\penalty0 (3):\penalty0 R46, 2014.

\bibitem[Wood et~al.(2019)Wood, Lu, and Langmead]{wood2019improved}
Derrick~E Wood, Jennifer Lu, and Ben Langmead.
\newblock Improved metagenomic analysis with kraken 2.
\newblock \emph{Genome biology}, 20:\penalty0 1--13, 2019.

\bibitem[Yu et~al.(2018{\natexlab{a}})Yu, Liu, Liu, Zhang, Magner, Lehnert,
  Qian, and Liu]{yu2018seqothello}
Y.~Yu, J.~Liu, X.~Liu, Y.~Zhang, E.~Magner, E.~Lehnert, C.~Qian, and J.~Liu.
\newblock Seqothello: querying rna-seq experiments at scale.
\newblock \emph{Genome Biology}, 19\penalty0 (1):\penalty0 167,
  2018{\natexlab{a}}.

\bibitem[Yu et~al.(2018{\natexlab{b}})Yu, Belazzougui, Qian, and
  Zhang]{yu2018memory}
Ye~Yu, Djamal Belazzougui, Chen Qian, and Qin Zhang.
\newblock Memory-efficient and ultra-fast network lookup and forwarding using
  othello hashing.
\newblock \emph{IEEE/ACM Transactions on Networking}, 26\penalty0 (3):\penalty0
  1151--1164, 2018{\natexlab{b}}.

\bibitem[Zentgraf et~al.(2020)Zentgraf, Timm, and Rahmann]{Zentgraf2020}
Jens Zentgraf, Henning Timm, and Sven Rahmann.
\newblock Cost-optimal assignment of elements in genome-scale multi-way
  bucketed cuckoo hash tables.
\newblock In Guy Blelloch and Irene Finocchi, editors, \emph{Proceedings of the
  Twenty-Second Workshop on Algorithm Engineering and Experiments (ALENEX)},
  pages 186--198. SIAM, 2020.

\end{thebibliography}

\end{document}